\documentclass[superscriptaddress, nofootinbib, preprintnumbers, twocolumn]{revtex4}

\usepackage{amsmath}	
\usepackage{amsthm}		
\usepackage{amssymb}	
\usepackage{eufrak}
\usepackage{datetime}
\usepackage{graphicx}
\usepackage{xcolor}
\usepackage{verbatim}
\usepackage{bm}
\usepackage{float}

\usepackage[colorlinks=true, citecolor=midblue, linkcolor=midblue, urlcolor=midblue]{hyperref}

\usepackage{setspace}

\definecolor{grey}{rgb}{0.4,0.4,0.4}
\definecolor{dullmagenta}{rgb}{0.4,0,0.4}
\definecolor{darkblue}{rgb}{0,0,0.4}
\definecolor{midblue}{rgb}{0,0,0.5}
\definecolor{midred}{rgb}{0.5,0,0}
\definecolor{orange}{rgb}{1,0.5,0}
\definecolor{lightbrown}{rgb}{0.75,0.5,0.25}
\definecolor{tan}{cmyk}{0.14,0.42,0.56,0}
\definecolor{djunglegreen}{cmyk}{0.99,0,0.52,0}
\definecolor{lightgreen}{rgb}{0,1,0}
\definecolor{olivegreen}{cmyk}{0.64,0,0.95,0.40}
\definecolor{midgreen}{rgb}{0.0,0.675,0.0}
\definecolor{darkgreen}{rgb}{0,0.5,0}

\newcommand{\vs}{\vspace}

\newcommand{\Irm}{\ensuremath{\mathrm{I}}}

\newcommand{\Prm}{\ensuremath{\mathrm{P}}}

\newcommand{\Urm}{\ensuremath{\mathrm{U}}}

\newcommand{\erm}{\ensuremath{\mathrm{e}}}

\newcommand{\urm}{\ensuremath{\mathrm{u}}}

\newcommand{\VEV}[1]{\langle #1 \rangle}

\newcommand{\QCD}{{\rm QCD}}
\newcommand{\PQ}{{\rm PQ}}

\newcommand{\osc}{{\rm osc}}

\newcommand{\const}{\mathrm{const}}

\newcommand{\eV}{\ensuremath{\, \mathrm{eV}}}

\newcommand{\GeV}{\ensuremath{\, \mathrm{GeV}}}
\newcommand{\TeV}{\ensuremath{\, \mathrm{TeV}}}

\smallskipamount

\settimeformat{ampmtime}

\usepackage{float}

\begin{document}

\title{Consequences of Multiple Axions in Theories with Dark Yang-Mills Groups}

\author{Manuel Ettengruber}
\email{manuel@mpp.mpg.de}
\affiliation{
	Arnold Sommerfeld Center,
	Ludwig-Maximilians-Universit{\"a}t,
	Theresienstra{\ss}e 37,
	80333 M{\"u}nchen,
	Germany,}
\affiliation{
	Max-Planck-Institut f{\"u}r Physik,
	F{\"o}hringer Ring 6,
	80805 M{\"u}nchen,
	Germany}
	
\author{Emmanouil Koutsangelas}
\email{emi@mpp.mpg.de}
\affiliation{
	Arnold Sommerfeld Center,
	Ludwig-Maximilians-Universit{\"a}t,
	Theresienstra{\ss}e 37,
	80333 M{\"u}nchen,
	Germany,}
\affiliation{
	Max-Planck-Institut f{\"u}r Physik,
	F{\"o}hringer Ring 6,
	80805 M{\"u}nchen,
	Germany}

\date{\formatdate{\day}{\month}{\year}, \currenttime}

\begin{abstract}

General consistency requirements of Quantum Gravity demand the existence of one axion per Yang-Mills group. In this work, we consider theories with dark Yang-Mills sectors and investigate general phenomenological implications of these necessary axions. We carry out computations for two simple models, namely a pure Yang-Mills sector and $N$ exact Standard Model copies. For the former, the misalignment mechanism results in a minimal dark confinement scale $\Lambda_{\rm conf} \gtrsim 1 \eV$ if the dark sector axion is supposed to make up the dark matter. For the latter, the misalignment mechanism without fine-tuning of the initial misalignment angle places an upper bound on $N$ below the species bound. When the PQ symmetries are broken during inflation, the collective isocurvature fluctuations do not necessarily tighten the bound on the inflationary Hubble scale arising from a single axion. We also point out that axion stars collectively made from axions of different dark sectors with a suppressed mass spectrum are not possible. Lastly, for the two models at hand, intersector interaction through axion kinetic mixing leads to the existence of two distinct axion states. For a single dark YM sector, the upper bound $\Lambda_{\rm conf} \lesssim 10^{12} \GeV$ emerges from the stability requirement of the dark sector axion. For $N$ exact SM copies, the mass and photon coupling of the second state is completely determined after a potential measurement of the analogous parameters of the first axion.

\end{abstract}

\maketitle

\tableofcontents

\section{Introduction}
\label{sec:Introduction}
\vs{-5mm}

Attempted measurements of the neutron electric dipole moment indicate no CP violation in the strong interaction. Since QCD includes CP violating processes proportional to a parameter $\theta$, the non-observation results in the bound $\theta < 10^{-10}$ \cite{ThetaBound}. This inexplicable smallness is referred to as the strong CP problem. While it looks like a naturalness problem this is actually misleading. It would be a naturalness problem if fine-tuning was necessary due to unacceptably large quantum corrections (like for the Higgs mass). For $\theta$ the quantum corrections from the Standard Model are much smaller than the measured bound \cite{ThetaCorrections}, so that from the point of view of naturalness it is rather a small value puzzle than a problem. 

However, the situation dramatically changes when quantum gravity is taken into account \cite{dvali2022strong}. In recent years, evidence has been provided that quantum gravity is incompatible with (meta-)stable de Sitter vacua \cite{Dvali:2013eja, Dvali:2018jhn}. In particular, this is organic in the S-matrix formulation of quantum gravity, which is only compatible with asymptotically flat space-times \cite{Dvali:2020etd}. This has fundamental implications for axion physics \cite{Dvali:2018dce, dvali2022strong}.  The point is that in its essence $\theta$ is not just a parameter but a label for QCD vacua. These vacua are non-degenerate. In particular, according to the Vafa-Witten theorem  \cite{PhysRevLett.53.535} the global minimum is the CP conserving one at $\theta = 0$. Other vacua have higher energies. Correspondingly, as argued in \cite{Dvali:2018dce, dvali2022strong}, if all theta vacua were to be physical, their non-degeneracy would imply the existence of de Sitter vacua among them. Moreover \cite{dvali2022strong}, even at the expense of
arbitrary fine-tuning, only one out of infinitely many vacua could be made viable for the S-matrix formulation. Others would end up in asymptotically non-flat cosmologies. Correspondingly, it was concluded in \cite{Dvali:2018dce, dvali2022strong} that
$\theta$ vacua must be unphysical in gravity. 

This means that when gravity is included the strong CP problem becomes a consistency problem and the non-degenerate vacuum structure must be removed, leaving only the viable vacuum. Therefore, a mechanism to select the right vacuum becomes indispensable \cite{Dvali:2018dce, dvali2022strong}, which was provided  by Peccei and Quinn (PQ) \cite{PQMechanism, PQMechanism2}. By introducing a non-linearly realized $\Urm(1)_\PQ$ that is anomalous with respect to QCD, the $\theta$-parameter is dynamically set to zero by the pseudo-Goldstone boson of $\Urm(1)_\PQ$ also known as the axion \cite{WeinbergAxion, Wilczek:1977pj}.

Viewed in this light, the strong CP problem is not exclusive to QCD but extends to any non-Abelian Yang-Mills (YM) group. Every YM group includes a vacuum angle that leads to an unacceptable de Sitter-type vacuum structure and must therefore be removed \cite{dvali2022strong}.

Using the PQ mechanism to achieve this, requires one axion per YM group because a single axion cannot set multiple $\theta$-parameters to zero. In particular, this is the case even when the different YM sectors are related by a discrete symmetry \cite{dvali2005threeform}, e.g. an exact permutation symmetry among $N$ Standard Model copies \cite{Dvali:2007iv}. This is due to the fact that the topological susceptibilities in different YM sectors break the discrete symmetry spontaneously. Correspondingly, a single axion can only eliminate one combination of the vacuum angles. The other combinations will remain physical and unconstrained, resulting in two problems. First, even ignoring gravity, this would imply that a single axion cannot solve the small $\theta$ puzzle for multiple YM groups \cite{dvali2005threeform, Dvali:2007iv}. Secondly, allowing these vacuum angles to persist, would necessarily introduce the asymptotically non-flat vacua that would conflict with the requirement of an S-matrix formulation, hence requiring one axion per YM group \cite{dvali2022strong}.

In this work, we consider theories in which dark matter includes one or more YM sectors and investigate the generic phenomenological consequences of the necessary axions. To our knowledge, axions have not been taken into account due to gravitational consistency reasons, which impose certain properties on them. 

To be precise, we consider models with $N$ sectors $\mathcal{L}_i$, each including at least one YM subgroup. Without loss of generality, we choose our sector to be labeled by $i=1$. All the others collectively make up the dark matter and we will refer to them as hidden or dark sectors. The sectors communicate via gravity and additional non-gravitational interactions, which we encode in $\mathcal{L}_{\rm mix}$. The associated Lagrangian can be written as 
    \begin{equation}
        \mathcal{L}
            =
                \sum_{i=1}^N \mathcal{L}_i
                + \mathcal{L}_{\rm mix}
            \; .
    \label{Eq:GeneralLagrangian}
    \end{equation}
The possible models can be divided in two categories. The first is a framework in which the dark sector consists of $N$ copies of the standard model related by an exact permutation symmetry \cite{Dvali:2009fw, Dvali:2009ne}. This case is most restrictive due to exact symmetry. Large values of $N$ are motivated by the Hierarchy problem \cite{Dvali:2007hz, Dvali:2007wp} but we shall also cover the case with smaller $N$. Such were discussed in different contexts, e.g., in \cite{Blinnikov:1982eh, Kolb:1985bf}.  

The second category includes all other possibilities. The dark sector in such cases is unrelated to the SM and can contain very different gauge structures. Such cases are expected to be rather generic in string theory compactifications \cite{PhysRevLett.54.502, Dixon:1985jw, Lebedev:2006kn, Braun:2005ux}. 

Before proceeding let us note that there exists a universal constraint on the extent of dark sectors in the form of the bound on the total number of particle species,  $N_s < 10^{32}$ \cite{Dvali:2007hz, Dvali:2007wp}. This bound is fully non-perturbative in its nature and originates from black hole physics, which implies that in theory with $N_s$ particle species the quantum gravity scale is lowered to $M_{*} = M_{P} / \sqrt{N_s}$ \cite{Dvali:2007hz}. From current phenomenology, this translates as the above bound on $N_s$, since in this extreme case the scale would be lowered to TeV. This would solve the hierarchy problem, which provides the main motivation for dark sectors with many species. Another important non-perturbative bound on the number of species originates from the constraints on the species entropy \cite{Dvali:2020wqi}. A more detailed discussion on both bounds on the number of species will be given later.

For concreteness, we will look at two simple models: One pure YM sector and $N$ exact SM copies. The former has been used to model dark matter due to the dynamics of YM theories, which for instance may result in a potential phase transition that produces stochastic gravitational waves during the early universe \cite{Halverson:2020xpg, PhysRevD.104.035005}. The latter was originally motivated by the many-species solution to the
hierarchy problem \cite{Dvali:2007hz, Dvali:2007wp}. In dealing with extended dark sectors, one can distinguish between universal effect and model-dependent features  characteristic of particular scenarios.  In particular, a universal feature is the modification of black hole physics at scales below the species scale $M_*$ (see,  \cite{Dvali:2008fd}). In addition, there are model dependent consequences. When applied to model the dark matter \cite{Dvali:2009fw}, it gives rise to many interesting phenomena such as neutron and neutrino oscillations \cite{Dvali:2009ne, Ettengruber:2022pxf}, modified black hole physics \cite{MicroBlackHolesDemocraticTransition}, and compact dark matter objects \cite{CompactDarkMatterObjectsViaN}. Even though we perform calculations for these two models, we shall try to keep our discussion maximally general and to point out the model independent predictions.

By adding to the dark YM sectors the necessary axions, we find several phenomenological consequences such as constraints on the parameter space of the dark sectors, potential new small structures, and new experimental signatures that we briefly summarize here.

The first consequence arises due to \textbf{axion production via the misalignment mechanism} \cite{MisalignmentPRESKILL, MisalignmentDINE, MisalignmentAbbott}. Today's axion energy density can of course be less than the observed dark matter density but it should not be higher. An important parameter is the initial misalignment angle of the axion field at the onset of the latest oscillations. As shown in \cite{Dvali:1995ce} this quantity is highly sensitive to earlier history and can be made arbitrarily small due to the early relaxation mechanism. We shall therefore treat it as a free input parameter for each axion sector. In the absence of an early relaxation mechanism \cite{Dvali:1995ce}, we find the following. For a single dark YM sector, this requirement puts a lower bound to the confinement scale of the dark YM sector, i.e. $\Lambda_{\rm conf} \gtrsim \eV$, if the dark sector axion is required to be the dominant contribution to the dark matter. In the case of more YM sectors the production potentially happens in several sectors. For $N$ exact SM copies, since each SM includes a QCD subgroup, this results in an upper bound on $N$. Notably, the misalignment mechanism is independent of the thermalization of the dark sectors and thus applies even for large values of $N$ when the sectors are very dilute. 

The presence of additional axions has another cosmological impact when the axions are present during inflation, massless during this time, and not driving inflation. Any axion that fulfills these conditions will be subject to \textbf{quantum fluctuations of isocurvature-type}. Since isocurvature fluctuations leave a special imprint in the CMB, their collective influence could tighten the bound on the inflationary Hubble scale arising from a single axion. We show that this is not necessarily the case, meaning that we identify regions in the parameter space where the isocurvature bound is avoided. 

In models with several dark sectors, it is possible that particles from different sectors collectively form structures \cite{CompactDarkMatterObjectsViaN}. While in \cite{CompactDarkMatterObjectsViaN} this phenomenon was studied with SM-like dark sectors, it does also apply to other compositions of the dark sectors. We show that this phenomenon can result in \textbf{boson stars that are collectively made from scalars of different dark sectors}. In accordance with the literature, we will call these $N$-boson stars. The interesting feature of such objects is that their size and mass are smaller by a factor of $1/\sqrt{N}$ compared to the single sector case. However, for axions, this phenomenon does not occur due to attractive self-interactions. 

Lastly, we ask the question if non-gravitational communication between the dark sectors can induce a modification of the axion physics in our sector. A potential source for this is \textbf{kinetic mixing between axions of different sectors}. We study this in the context of $N$ exact SM copies where the underlying discrete symmetry forces all kinetic mixing parameters to be equal. We argue that kinetic mixing emerges in both, the standard PQ scenario as well as in the two-form gauge implementation of the axion, and that in the standard PQ scenario, the parameters encoding the strength of the kinetic mixing, $\epsilon$, are suppressed by $1/N$ due to perturbative unitarity. The kinetic mixing results in $N-1$ degenerate states and one special state, which is lighter by a factor of $1/\sqrt{2}$ and has couplings suppressed by $1/\sqrt{N}$ (calculated for large values of $N$) as compared to the heavier axion states. Interestingly, the lighter state is completely determined by the heavier states, meaning that its mass and couplings are predicted once the mass and couplings of the heavier states are known. For a single dark YM sector, the mixing induces a decay channel of the dark sector axion into photons. If the dark sector axion is supposed to make up the dark matter, its stability requirement results in an upper bound on the dark confinement scale of $\Lambda_{\rm conf} \lesssim 10^{12} \GeV$.

It is also worth mentioning that in the case of many dark sectors including their associated axions, we arrive at a situation that is reminiscent of the string theory inspired axiverse \cite{Arvanitaki:2009fg}. Interestingly, the axions in our framework arise from a motivation that is completely different from the one in string theory. The phenomenology of several axions in the context of the axiverse was for instance studied in \cite{Mack:2009hs}. 

The paper is organized as follows. To begin with, in Sec.~\ref{sec:Cosmological_Implication} we start by reviewing the cosmological framework of $N$ dark sectors. We then discuss how to implement the PQ mechanism in each sector and show the bounds arising from the misalignment production and isocurvature perturbations of $N$ axions. Lastly, we discuss collective bound states of multiple axions. In Sec.~\ref{sec:Kinetic_Mixing} we study the consequences of kinetic mixing between $N$ axions and point out the differences for different types of experiments. Finally, in Sec.~\ref{sec:Summary_and_Outlook} we summarize our results and give an outlook.

\section{Cosmological Implications}
\label{sec:Cosmological_Implication}
\vs{-5mm}

\subsection{Cosmological Framework}
\label{sec:Cosmological_Framework}
\vs{-5mm}

To begin with, let us clarify the behaviour of dark sectors within a cosmological framework. Let us, for now, neglect intersector interactions except gravity by setting $\mathcal{L}_{\rm mix} = 0$ and discuss them separately in Sec.~\ref{sec:Kinetic_Mixing}. This will not affect the conclusions of this section.

For the viability of these models in a cosmological framework, the dark sectors must behave as a cold and pressureless fluid that forms stable dark matter halos. In addition, if the dark sectors include any massless fields, they need to avoid the bound on the number of massless particles during nucleosynthesis. These conditions might seem difficult to realize, especially in the presence of non-trivial interactions in the dark sectors. We will apply the solution proposed in \cite{Dvali:2009fw} to achieve this. 

There, inflation was used to populate and reheat the sectors differently. The crucial point is that the reheating into the dark sectors is far less efficient than in the visible sector. The consequent low reheating temperature in the dark sectors is used to avoid the nucleosynthesis bound, suppress cooling processes that result in the collapse of dark matter halos, and generally result in a different behavior throughout the cosmological history despite the potential similarity to the visible sector. For large $N$, when each sector is so dilute that it is problematic to regard them as thermalized at all, the behaviour as a pressureless fluid arises from kinematics, i.e. from the fact that particles from the same sector barely meet each other.

The situation is unchanged even when starting with equal sectors. Cosmological evolution breaks the symmetry in a sense by resulting in different energy densities and temperatures. The field theoretic parameters of each sector, however, remain equal.

\subsection{Bounds on the number of species}

Before introducing axions in each sector, let us discuss the constraints on N. These constraints arise from the bounds on the number of species $N_s$ due to their interactions with gravity and via other forces. First, there exists a gravitational bound on the number of particle species imposed by black hole physics, which tells us that in a theory with $N_s$ particle species the gravitational cutoff is not given by the Planck mass, but by the following scale (usually called the species scale) \cite{Dvali:2007hz, Dvali:2007wp},

\begin{equation}
  M_* = \frac{M_P}{\sqrt{N_s}}
  \; .
  \label{fundamentalscaleofgravity}
\end{equation}
The scale $1/M_*$ gives the size of the smallest possible black hole. 
\par 
Secondly, there exists another universal bound on species independent of gravity. This bound says that the number of species $N_s$ inter-coupled by a coupling
of strength $\alpha(q)$ and the scale (momentum transfer) $q$ is bounded by \cite{Dvali:2020wqi, Dvali:2021jto}
\begin{equation}
    N_s = \frac{1}{\alpha(q)}
    \; ,
    \label{speciescouping}
\end{equation}
where the coupling $\alpha$ must be evaluated at the scale $q$. As shown in \cite{Dvali:2020wqi} this bound originates from species entropy since for a larger number of species the unitarity would be violated due to multi-particle states with exceedingly high entropy of species. In general, as shown there, no state in the theory can have an entropy exceeding the value given by \eqref{speciescouping}. The coupling entering in the bound counts the subset of species that are intercoupled. If the species are intercoupled selectively according to some rules, the bound is modified accordingly. In each case, the guideline is the entropy of species which leads to the conclusion that the coupling should not exceed the bound given in \eqref{speciescouping}. For explicit examples, see, \cite{Dvali:2020wqi}. For cosmological implications of the species entropy bound see \cite{Dvali:2023xfz}. Also, as explained there, from this perspective the black hole bound \eqref{fundamentalscaleofgravity} on species represents a particular case of species entropy bound \eqref{speciescouping}, since the quantity $(M_*/M_P)^2$ represents a gravitational coupling evaluated at the scale $p = M_*$. 
\par 

In our analysis the minimal case would be a single extra axion ($N=2$ in total) which corresponds to a single gauge group  group in dark sector.  In case when the dark sector is represented by the exact SM copies, the minimal number would be a single mirror copy. Possibility of dark matter in such a copy was studied e.g., in \cite{Blinnikov:1982eh, Kolb:1985bf}. However, this value is excluded if the dark copy is supposed to make up all the dark matter since in this case, the dark matter halos would have collapsed into substructures. By increasing the number of SM copies, the particle density in each sector is reduced, leading to a suppressed dissipation efficiency and the lowest phenomenologically viable value of $N$ being of $\mathcal{O}(10)$ \cite{Dvali:2007wp, Dvali:2020wqi}.

\subsection{The Axion Mechanism}
\label{sec:The_PQ_Mechanism}
\vs{-5mm}

In order to eliminate the $\theta$-vacua let us now ad an axion \cite{WeinbergAxion, Wilczek:1977pj} to each gauge sector.  First, this can be done via a traditional PQ formulation \cite{PQMechanism, PQMechanism2}, which requires a spontaneously broken global $U(1)_{PQ}$ symmetry.   
This means we introduce a global U(1)$_\PQ$ in each sector that is chiral, non-linearly realised and anomalous with respect to the respective YM group. In the low-energy effective theory the axion fields $a_i(x)$ are described as pseudo-scalars with the anomalous couplings 
    \begin{equation}
        \mathcal{L}^{\rm anomalous}_i
            \propto
                \left(
                    \frac{a_i}{f^{a_i}} + \theta_i
                \right)
                G_i \Tilde{G}_i
            \; ,
    \end{equation}
where the $G^i$ denote the YM field strength tensors, $\Tilde{G}^i$ their duals, and $f^{a_i}$ the axion decay constants. Below the confinement scale, these couplings transform into potentials for the axions, which according to the Vafa-Witten theorem have their vacua at $\langle a_i \rangle = - f^{a_i} \theta_i$. Expanding around these vacua removes the vacuum angles from the theory. Note that in more complex theories each YM vacuum angle receives additional contributions, e.g. from the phases of the quark matrices in the SM. We do not introduce an additional symbol for this and keep the different contributions implicitly encoded in $\theta_i$.

The implementation of the U(1)$_\PQ$ with the required properties into the SM is usually achieved by the KSVZ and DFSZ invisible axion models \cite{DFSZ1, DFSZ2, KSVZ1, KSVZ2}. The KSVZ model introduces an SM scalar singlet together with a heavy quark, while the DFSZ model uses a singlet and an additional Higgs doublet. In the former, the singlet $\Phi_1$ acquires a vacuum expectation value $f^{a_1}$ via a standard Mexican hat potential,
    \begin{equation}
        V_1(\Phi_1)
            =
                \lambda_1
                \left( 
                    |\Phi_1|^2 - (f^{a_1})^2 
                \right)^2
            \; 
    \end{equation}
which generates the spontaneous breaking of the U(1)$_\PQ$. The axion is then nothing else than the angular degree of freedom of the singlet, which we canonically normalize as $\theta_1(x) \equiv a_1(x)/f^{a_1}$. In the SM the introduction of the singlet is necessary to decouple the electroweak scale from the PQ scale $f^{a_1}$, thus avoiding star cooling bounds and direct detection until now. Here we focus on the minimal invisible axion models but in principle, there exists a plethora of KSVZ- and DFSZ-type models, which could be realized in our sector (see \cite{DiLuzio:2017pfr, Plakkot:2021xyx} and \cite{DFSZ} for catalogs of preferred models in each category).

The massless quark solution to the strong CP problem is in fact also a realization of the PQ mechanism. In the presence of a massless quark, say the up-quark in the SM, U(1)$_\PQ$ symmetry coincides with the U(1)$_A$ symmetry of QCD and the $\eta'$ meson plays the role of the axion \cite{dvali2005threeform}. This implementation is not realized in the SM since all SM quarks are massive, but it can be realized in a dark sector.

However, the axion mechanism can alternatively be realized via an intrinsic gauge formulation, originally introduced in \cite{dvali2005threeform} and deeper investigated in\cite{dvali2022strong} (see also \cite{PhysRevD.105.085020} for a consistency check). This formulation does not involve any global $U(1)_{PQ}$ symmetry. Instead, it is based on gauge redundancy of the respective QCD like theory. Thus, in gauge formulation, the axion is introduced as a two-form $B^1_{\mu\nu}$ that becomes the longitudinal mode of the QCD Chern-Simons three-form $C^1_{\mu\nu\rho}$ (similar to the scalar that becomes the longitudinal mode of the Maxwell-field in the Stückelberg formulation of the Proca theory). This fixes the form of the theory, to the lowest order, as, 
	\begin{equation}	
        \mathcal{L}_{\rm eff}
			=  
				- \frac{1}{2 \cdot 4!}
				(F^1_{\mu \nu \rho \sigma})^2
				+ \frac{m^2}{2 \cdot 3!} 
				\left(
					C^1_{\mu\nu\rho} 
					- \partial_{[\mu} B^1_{\nu\rho]}
				\right)^2
			\; ,
	\end{equation}
where $F^1_{\mu \nu \rho \sigma} \equiv \partial_{[\mu} C^1_{\nu\rho\sigma]}$ denotes the canonically normalized field strength tensor of the Chern-Simons three-form. This puts the Chern-Simons theory into the Higgs phase, effectively rendering the vacuum angle unphysical. The clear advantage of this implementation is that it does not suffer from the quality problem, meaning that $\theta_1 = 0$ can not be spoiled by any explicit breaking, since no global symmetry is required. Correspondingly, the mechanism is stable under any gauge invariant deformation of the theory.  As discussed in \cite{dvali2022strong}, one of the consequences of this realization is that any measurement of a neutron electric dipole moment would imply new physics, in contrast to the ordinary PQ implementation where a non trivial $\theta_1$ could come from any explicit breaking. Below $f^{a_1}$ the pseudo-scalar $a_1(x)$ and the two-form axion $B^1_{\mu\nu}$ are dual to each other so that the physics is vastly the same apart from the potential explicit breaking in ordinary PQ realization. Above $f^{a_1}$, however, the formulations are very different and the gauge formulation likely requires its UV-completion directly in quantum gravity. This implies a PQ scale of $f^{a_1} \sim M_\Prm$ (or $f^{a_1} \sim M_*$ in the presence of a large number of species). Obviously, the gauge axion realization can be implemented for arbitrary hidden sector gauge groups.

In the dark sectors, the concrete implementation of the U(1)$_\PQ$ depends on the model but in principle, there is more freedom due to a lack of bounds. For exact SM copies, we could again use the KSVZ or DFSZ models or the two-form implementation. 
For pure YM sectors, while we could again implement the axion mechanism by using the KSVZ scenario or the gauge two-form implementation, the massless quark solution is also viable.

In general early cosmologies of  gauge and $U(1)_{PQ}$  axions can be very different, in particular, due to fundamental differences in their UV completions above the scale $f_a$. However, our initial conditions  make these potential differences unimportant. Hence, we shall remain independent of any concrete realization of the axion mechanism and treat $f^{a_i}$ as free parameters. However, if necessary, we will discuss consequences that are specific to certain realizations.

\subsection{Misalignment: Exact SM Copies}
\label{sec:Exact SM Copies}
\vs{-5mm}

In the single axion case a very stringent bound on the axion scale results from the misalignment mechanism being required not to produce more dark matter than observed \cite{MisalignmentPRESKILL, MisalignmentDINE, MisalignmentAbbott}. An important input quantity in this bound is the initial misalignment angle $\theta^{\rm ini}$, which can be defined as the value in the latest epoch when the Hubble parameter crosses below the axion mass. Naturally, the most stringent bound is obtained for $\theta^{\rm ini} \sim \mathcal{O}(1)$. However, as shown in \cite{Dvali:1995ce} this value is highly sensitive to the details of the previous evolution of the axion field (e.g., to the value of the QCD scale during inflation) and can be made arbitrarily small via the early relaxation mechanism. Due to this, in our analysis, we shall treat $\theta^{\rm ini}$ in each gauge sector as independent free parameters. Under these circumstances, we shall derive constraints on the parameter space for models with multiple axions. 

For a clearer presentation, let us begin with the case of $N$ exact SM copies, where the equality between the sectors can be guaranteed by a discrete symmetry \cite{Dvali:2007hz, Dvali:2007wp}. Various phenomenological aspects of this framework were studied in \cite{FOOT199167,Dvali:2009ne}. As already explained \cite{Dvali:2007iv}, despite the exact permutation symmetry of the Lagrangian, the solution of the small $\theta$ puzzle requires the existence of $N$ axions. The same is demanded by the S-matrix consistency
of gravity \cite{dvali2022strong}. This necessitates the existence of $N$ axions. For alternative compositions of the dark sectors we then only discuss the differences in the calculation and give the final result. 

In the standard axion picture, the U(1)$_\PQ$ symmetry is spontaneously broken at $T^\PQ \sim f_a$. In our framework, this takes place in every sector at the temperatures $T^\PQ_i \sim f_{a_i} \equiv f_a$. The axions $\theta_i(x)$ emerge as the corresponding Goldstone bosons and, as such, have flat potentials. Using the standard notation we canonically normalize the Goldstone-fields as $\theta_i(x) \equiv a_i(x)/f_a$. 

The potential of each axion is flat only when the corresponding QCD is in the Coulomb phase. In the confinement phase, each axion receives an effective potential from non-perturbative effects of its corresponding QCD, which to leading order in the semi-classical approximation can be accounted by instantons. In a cosmological setting, these instantons are coupled to the thermal bath, which causes the axion masses to depend on temperature. For instance, in the dilute instanton gas approximation \cite{PhysRevD.17.2717}, the potential takes the form \cite{QCDINstantonsFiniteTemp}
    \begin{equation}
        V_i(\theta_i)	
            =
                m_a^2(T_i) f_a^2
                \big(1 - \cos(\theta_i)\big)
            \; ,
    \label{Eq:PotDIGA}
    \end{equation}
where the Temperature-dependent masses are given by
    \begin{equation}
        m_a(T_i)
            \equiv
                \frac{(\Lambda_\QCD^3 m_u)^{\frac{1}{2}}}{f_a}
                \begin{cases}
                    \beta \,
                    \left(
                        \frac{\Lambda_\QCD}{T_i}
                    \right)^4 
                    &:T_i > \Lambda_\QCD
                \; ,
                \\
                    1 
                    &:T_i \lesssim \Lambda_\QCD
                \; .
                \end{cases}
    \label{Eq:AxionMassInstanonThermal}
    \end{equation}
Here, $m_u$ denotes the up-quark mass while $\beta$ encodes QCD and active quark physics at the temperature $T_i$. For the SM, we roughly have $\beta \sim 10^{-2}$ \cite{PhysRevD.17.2717}. 


In an FLRW background with $R(t)$ denote the scale factor and $H(t)$ the Hubble parameter, the equation of motion for each axion takes the form
    \begin{equation}
        \ddot{\theta}_i 
        + 3H(t) \dot{\theta}_i 
        - \frac{1}{R^2(t)} \Delta \theta _i
        + \frac{V_i'(\theta_i)}{f_a^2} 
            = 
                0
            \; .
    \label{Eq:EomTheta}
    \end{equation}
Let us use the potential \eqref{Eq:PotDIGA} and make the following two simplifications. First, only the leading order in $\theta_i$ is considered in the potential. Secondly, only the zero mode of $\theta_i$ is taken into account. With these simplifications, each equation of motion reduces to that of a damped harmonic oscillator,
    \begin{equation}
        \ddot{\theta} _i
        + 3H(t) \dot{\theta} _i
        + m_a^2\big( T_i(t) \big) \theta_i
            = 
                0
            \; .
    \label{Eq:KGTheta}
    \end{equation}

At $T_i >> \Lambda_\QCD$ the potential is flat so that $\theta_i \approx \const$, i.e. the i-th axion is basically frozen out. At $T_i \sim \Lambda_\QCD$ the corresponding axion potential gets significant but it is not until the Hubble friction is overcome before the axion starts performing coherent oscillations around the vacuum. We define this critical time via 
    \begin{equation}
            m_a \big( T_i(t_\osc) \big) 
                = 
                    3 H(t_\osc)
                \; .
    \label{Eq:OscCondition}
    \end{equation}
From that moment on its equation of state no longer correspond to that of dark energy but to that of non-relativistic matter, making the corresponding axion contribute to the dark matter energy density. 

For our model, this means that, due to the very low temperature in the dark sectors, the axions there are essentially created with their zero-temperature potential switched on. In contrast, our axion is created with a flat potential. Therefore, since the zero-temperature mass is larger than the high-temperature mass, the axion from our sector will start oscillating later than the other axions. In order to quantify this, let us use our sector's temperature $T_1 \equiv T$ as a clock instead of the cosmic time $t$. In terms of $T$ the Hubble parameter during radiation domination is given by
    \begin{equation}
        H (T)
            =
                \sqrt{\frac{\rho_{\rm tot}}{3 M_\Prm^2}}
            \sim
                    \frac{ T^2 }
                    {M_\Prm}
            \; .
    \label{Eq:HubbleT1}
    \end{equation}
Here, $\rho_{\rm tot}$ denotes the total energy density, which is dominated by our sector. Using (\ref{Eq:HubbleT1}) and the axion masses defined by (\ref{Eq:AxionMassInstanonThermal}), the condition (\ref{Eq:OscCondition}) results in the oscillations commencing when
    \begin{align} \nonumber
        T_{\osc ,1}
            &\sim
                \left[
                    \frac{
                        \beta M_\Prm\Lambda_\QCD^\frac{11}{2} m_u^\frac{1}{2}}{3f_a}
                \right]^\frac{1}{6} \\
            &\sim 
                    4 \times 10^{-1} \GeV \left(
                        \frac{10^{12} \GeV}{f_a}
                    \right)^\frac{1}{6}
            \; ,
    \label{Eq:TOSC1}
    \end{align}
in our sector and
    \begin{align} \nonumber
        T_{\osc ,i}
            &\sim
                \left[
                    \frac{M_\Prm\Lambda_\QCD^\frac{3}2 m_u^\frac{1}{2}} {3f_a}
                \right]^\frac{1}{2} \\
            &\sim 
                    2 \times 10^{1} \GeV \left(
                        \frac{10^{12} \GeV}{f_a}
                    \right)^\frac{1}{2}
            \; ,
    \label{Eq:TOSC2}
    \end{align}
in the dark sectors with $i\neq 1$. Note that both moments in time are expressed in terms of our sector's temperature. 

While we left $f_a$ as a free parameter there is a small caveat for $f_a \gtrsim 6 \times 10^{17} \GeV$. In this range our axion has also reached its zero-temperature mass before overcoming the Hubble friction \cite{Fox:2004kb}, so that for these values of $f_a$ all axions start oscillating at the temperature dictated by (\ref{Eq:TOSC2}).

The initial energy density of each oscillation is
    \begin{equation}
        \rho_{a_i}(T_{\osc ,i})
            = 
                \frac{1}{2} f_a^2 
                m_{a_i}^2(T_{\osc ,i}) \theta^2_i(T_{\osc ,i})
            \; ,
    \end{equation}
where $\theta_i(T_{\osc ,i}) \equiv \theta_i^{\rm ini}$ is each sectors initial misalignment angle. Essentially, we define the initial misalignment angle $\theta_i^{\rm ini}$ as the value when the axion mass crosses below the Hubble parameter in most recent history. However, as shown in \cite{Dvali:1995ce} this quantity is strongly sensitive to the pre-history and can be made arbitrarily small due to early relaxation mechanisms. We thereby shall treat $\theta_i^{\rm ini}$ as a free parameter.

Since Big Bang Nucleosynthesis took place during radiation domination, the dark axions must not dominate the energy density of the universe at $T_{\osc,i}$. The energy density for relativistic degrees of freedom at that time is given by $\rho_{\rm rad}(T_{\osc,i}) \sim g_*(T_{\osc,i}) (T_{\osc,i})^4$, where the effective number of relativistic species is $g_*(T_{\osc,i}) \sim 10^2$. The requirement $\sum_{i=2}^N \rho_{a_i}(T_{\osc,i}) \ll \rho_{\rm rad}(T_{\osc,i})$ results in the bound
    \begin{equation}
        N
            \lesssim
                \frac{g_*(T_{\osc,i}) T_{\osc,i}^4}
                {\Lambda_\QCD^3 m_\urm (\theta_i^{\rm ini})^2}
            \sim 
                10^{12} 
                 \left(
                    \frac{10^{12} \GeV}{f_a}
                \right)^2
                \left(
                    \frac{1}{\theta_i^{\rm ini}}
                \right)^2
            \; .
    \end{equation}

As it turns out, a stronger bound on $N$ can be found at today's Temperature. Since the axions are decoupled, the number of zero modes per co-moving volume is conserved as long as the changes in the mass are in the adiabatic regime. Assuming this to be the case in every sector, today's energy density per sector is
    \begin{equation}
        \rho_{a_i}(T_{\rm today})
            = 
                \rho_{a_i}(T_{\osc,i})
                \frac{m_{a_i}(T_{\rm today})}{m_{a_i}(T_{\osc,i})}
                \left(
                    \frac{T_{\rm today}}{T_{\osc,i}}
                \right)^3
            \; 
    \end{equation}
We normalize with respect to the critical energy density $\rho_{\rm cr} \sim M_\Prm^2 H_{\rm today}^2$ to receive the corresponding axion fraction $\Omega_{a_i}$ in today's universe. By comparing $\Omega_{a_i}$ with the dark matter fraction $\Omega_{\rm DM}$ from the latest Planck mission \cite{Aghanim:2018eyx}, we find for our sector
    \begin{equation}
        \frac{\Omega_{a_1}}{\Omega_{\rm DM}}
            \sim 
                0.54 
                \left(
                    \frac{\beta^{-\frac{1}{6}} m_u^\frac{5}{12} \Lambda_\QCD^\frac{7}{12}}{10^{-2}\GeV}
                \right)
                \left(
                    \frac{f_a}{10^{12}\GeV}
                \right)^\frac{7}{6}
                \left(
                    \frac{\theta_1^{\rm ini}}{1}
                \right)^2                
            \; ,  
    \label{Eq:DMfraction1}
    \end{equation}
while for each dark sector, i.e. $i\neq 1$, we get
    \begin{equation}
        \frac{\Omega_{a_i}}{\Omega_{\rm DM}}
            \sim 
                0.01
                \left(
                    \frac{m_u^\frac{1}{4}\Lambda_\QCD^\frac{3}{4}}{10^{-2}\GeV}
                \right)
                \left(
                    \frac{f_a}{10^{12}\GeV}
                \right)^\frac{3}{2}
                \left(
                    \frac{\theta_i^{\rm ini}}{1}
                \right)^2                                
            \; .  
    \label{Eq:DMfractioni}
    \end{equation}
In summary, the variations in reheating temperatures cause the axion oscillations to commence at different times, resulting in varying axion densities despite identical field theoretic parameters. While a single mirror sector has an insignificant cosmological influence due to lower axion density compared to our axion, larger numbers of mirror sectors result in the accumulation of densities and their collective effect cannot be ignored. In the following, we will examine the parameter space and explore the consequences of this collective effect.

For our axion, a lower bound follows from astrophysics due to non-trivial couplings to matter \cite{Raffelt:2006cw}. These couplings lead to a contributions in the cooling of stars that result in the bound $f_a \gtrsim 10^9 \GeV$. Although other axions are not coupled to our sector directly and are thus not contributing to the cooling in stars, the PQ scale in all sectors must fulfill this bound due to the exact permutation symmetry between the sectors.

\begin{figure}[t]
    \centering
    \includegraphics[width=0.483\textwidth]{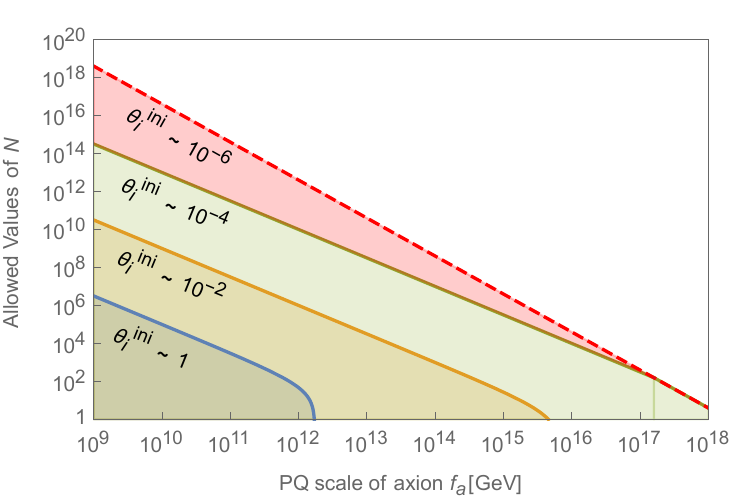}
    \caption{The allowed parameter space with different initial misalignment angles $\theta_i^{\rm ini}$ according to (\ref{Eq:MisalignmentBoundonN}). The values on the thick lines correspond to the case where the dark matter is entirely composed of axions. The dashed line presents the species bound, meaning that along this line the gravitational cutoff $M_*$ and $f_a$ coincide.}
    \label{fig:RegionN}
\end{figure}

\begin{figure}[t]
    \centering
    \includegraphics[width=0.483\textwidth]{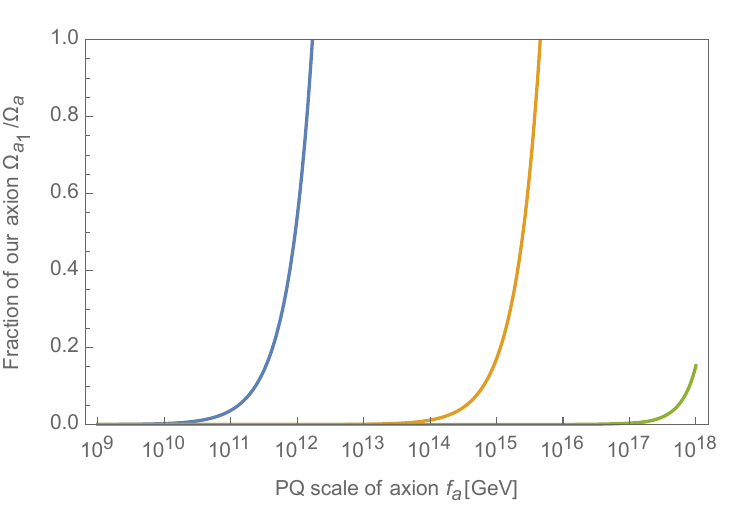}
    \caption{The fraction of our axion energy density with respect to the total axion energy density in the case when the axions make up all the dark matter. For larger $f_a$ our axion is dominating, while for lower $f_a$ the axions from dark sectors dominate.}
    \label{fig:OurAxionFraction}
\end{figure}

While our axion density matter is not allowed to exceed the observed dark matter density, leading to the known bound $f_a \lesssim 10^{12} \GeV$, the total axion density must not exceed it as well, i.e.
    \begin{equation}
        \sum_{i=1}^N \frac{\Omega_{a_i}}{\Omega_{\rm DM}}
            \lesssim 
                1
            \; .
    \label{Eq:AxionConstrainReq}
    \end{equation}
This results in the following inequality for the viable parameter space, 
    \begin{align}\nonumber
        N
           \lesssim 
                & 10^2
                \left(
                    \frac{10^{12} \GeV}{f_a}
                \right)^{3/2}
                \left(
                    \frac{1}{\theta_i^{\rm ini}}
                \right)^2 \\
                & \times \left[
                    1 - 0.54 
                    \left(
                       \frac{f_a}{10^{12}\GeV}
                    \right)^\frac{7}{6}
                    \left(
                        \frac{\theta_1^{\rm ini}}{1}
                    \right)^2  
                \right]
            \; .
    \label{Eq:MisalignmentBoundonN}
    \end{align}
We depict the viable regions in the $N$ - $f_a$ - plane for different values of $\theta_i^{\rm ini}$ in Figure \ref{fig:RegionN}. The thick lines indicate the combinations that result in the axions making up all the dark matter, corresponding to the equality in (\ref{Eq:AxionConstrainReq}). For example along the blue line in the plot, the entire dark matter is composed of our axion when $f_a \sim 10^{12} \GeV$, whereas for $f_a \sim 10^{9} \GeV$ essentially the $N \sim 10^6$ axions from the dark sectors make up the dark matter. A similar behaviour is observed for the other depicted values of $\theta_i^{\rm ini}$.

We can quantify this by calculating the fraction of our axion along these lines, which we depict in Figure~\ref{fig:OurAxionFraction}. There we also see that for a fixed $\theta_i^{\rm ini}$, lower $f_a$ values result in a dominant contribution from dark sector axions, while larger $f_a$ values lead to a dominant contribution from our axion.

Let us briefly discuss the range of the initial misalignment angles. The value of $\theta_i^{\rm ini}$ depends on whether the PQ symmetry is broken during or after inflation. When it is broken after inflation, for each sector there appear all possible values of $\theta_i^{\rm ini}$ in today's Hubble patch so that one can average over the uniform distribution resulting in $\theta_i^{\rm ini} = \pi/\sqrt{3}$ for all $i$. When the PQ symmetry is broken during inflation, the initial misalignment angle is in principle an arbitrary initial condition and $\theta_i^{\rm ini} \sim \mathcal{O}(1)$ seems to be a reasonable choice in the absence of an explanation for special initial conditions. However, such initial conditions appear naturally when QCD becomes strong during inflation, yielding $\theta_i^{\rm ini} \ll 1$ \cite{Dvali:1995ce} (see also \cite{Takahashi:2018tdu, Koutsangelas:2022lte} for concrete implementations). If such a phase or a similar mechanism existed in each sector, the parameter space would extend to $f_a \gg 10^{12}\GeV$ and $N \gg 10^6$. Note that for a given $N$, the PQ scale is bounded from above by (\ref{Eq:SpeciesBound}), which we depict via a red dashed line in Figure~\ref{fig:RegionN}. 

The existence of such a phase in every sector would allow for $N > 10^6$ but this poses the question if this is enough to solve the Hierarchy problem. As mentioned in the previous section, the existence of $N \sim 10^{32}$ copies of the SM would provide a solution to the hierarchy problem. As can be read of from Figure~\ref{fig:RegionN} or equivalently from (\ref{Eq:MisalignmentBoundonN}), this large number requires $f_a \ll 10^9 \GeV$ together with an early phase of strong QCD. This is in conflict with the lower bound on the PQ scale from astrophysical considerations in our sector, i.e. $f_a \gtrsim 10^9\GeV$. Seemingly this excludes a large number of exact SM copies as a solution for the hierarchy problem. However, there are proposals such as the clockwork mechanism \cite{Kaplan:2015fuy} that allow the separation of the PQ scale from the suppression factor of the couplings. This would render the part of the parameter space that solves the hierarchy problem viable. 

\subsection{Misalignment: Pure YM Dark Sector}
\label{sec:Pure_YM}
\vs{-5mm}

Finally, let us discuss the case where each dark sector is based on a pure YM group SU$_i(N_C)$. The crucial difference compared to the case with exact SM copies would be the absence of light quarks. Without light quarks, chiral perturbation techniques can no longer be used to calculate the zero-temperature mass of the axion. Instead, the mass would have to be calculated by using alternatives such as large $N_c$ methods. We use as an approximation for the axion mass the extrapolated result of the dilute instanton gas at finite temperatures \cite{QCDINstantonsFiniteTemp}, 
    \begin{equation}
        m_{a_i}(T^i)
            \equiv
                \frac{(\Lambda^i_{\rm conf})^2}{f_{a_i}}
                \begin{cases}
                    \left(
                        \frac{\Lambda^i_{\rm conf}}{T^i}
                    \right)^4 
                    &:T^i > \Lambda^i_{\rm conf}
                \; ,
                \\
                    1 
                    &:T^i \lesssim \Lambda^i_{\rm conf}
                \; ,
                \end{cases}
    \label{Eq:AxionMassInstanonThermal2}
    \end{equation}
where $\Lambda_{{\rm conf},i}$ denotes the different confinement scales of the dark sectors and the factor of $\beta$ was dropped due to the absence of light quarks. In contrast to the case of exact SM copies, the $\Lambda_{{\rm conf},i}$ are free parameters that in principle can be even smaller than the dark sector temperatures. Hence, the dark sector axion can in principle be created with an essentially flat potential, although this requires quite low confining temperatures. In the following, we will only consider the case where the dark axions are created with their zero-temperature mass.

\begin{figure}[t]
    \centering
    \includegraphics[width=0.483\textwidth]{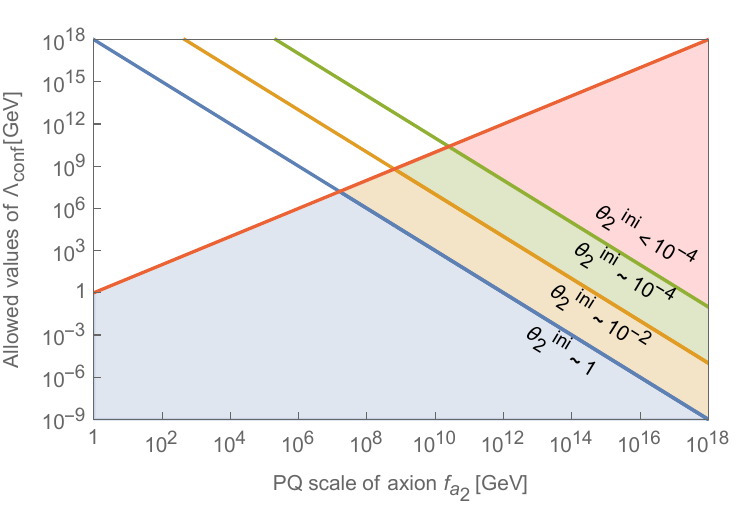}
    \caption{The allowed region for a single dark, pure YM sector with confinement scale $\Lambda_{\rm conf}$, axion scale $f_{a_2}$, and no intersector interactions. Minimality favors $\theta^{\rm ini}_2 \sim 1$, since in this scenario there is no need for additional physics that results in a small misalignment angle. The red line represents the perturbative unitarity bound $\Lambda_{\rm conf} \lesssim f_{a_2}$.}
    \label{fig:Lambdai}
\end{figure}

From there the calculation is vastly the same as in the exact SM case. The oscillations in the dark sectors commence at
    \begin{align} \nonumber
    T_{\osc ,i}
        \sim
            \Lambda^i_{\rm conf}
            \left(
                \frac{M_\Prm}{3f_{a_i}}
            \right)^\frac{1}{2}
        \; ,
    \label{Eq:TOSC2PureYM}
    \end{align}
which results in the final dark sector density of
    \begin{equation}
        \frac{\Omega_{a_i}}{\Omega_{\rm DM}}
            \sim 
                10^{-18}
                \left(
                    \frac{\Lambda^i_{\rm conf}}{\GeV}
                \right)
                \left(
                    \frac{\theta_i^{\rm ini}}{1}
                \right)^2  
                \left(
                    \frac{f_{a_i}}{\GeV}
                \right)^\frac{3}{2}
            \; .
    \label{Eq:DMfractioniPureYM}
    \end{equation}

Let us discuss the parameter space for the particularly interesting case of a single, pure YM sector, i.e. $N=2$ with $i=2$ being the dark sector. To simplify the notation, we define $\Lambda^{i=2}_{\rm conf} \equiv \Lambda_{\rm conf}$. For the dark sector axion to not result in an abundance of dark matter, we find
    \begin{equation}
        \Lambda_{\rm conf}
            \lesssim
                10^{18}\GeV
                \left(
                    \frac{1}{\theta_2^{\rm ini}}
                \right)^2  
                \left(
                    \frac{\GeV}{f_{a_2}}
                \right)^\frac{3}{2}
            \; ,
            \label{Eq:Confinementupperbound}
    \end{equation}
where the equality is valid when the dark axion makes up all the dark matter. Furthermore, to ensure perturbative unitarity in the dark sector, we must have 
    \begin{equation}
        \Lambda_{\rm conf}
            \lesssim   
                f_{a_2}
            \; .
    \end{equation}
This follows from the requirement of the axion's quartic self-interaction to be in the weak coupling regime. We depict the viable region in Figure~\ref{fig:Lambdai}. If the dark sector axion is supposed to make up the dark matter, we see that for $\Lambda_{\rm conf}$ to lie below the Planck scale, the dark sector PQ scale must fulfill $f_{a_2}\gtrsim 1 \GeV$, which is pushed higher for smaller values of $\theta_2^{\rm ini}$. Equivalently, for $f_{a_2}$ to lie below the Planck scale, the dark confinement scale must fulfill $\Lambda_{\rm conf} \gtrsim 1 \eV$. This minimum value is also pushed higher for smaller values of $\theta_2^{\rm ini}$. 

It should be mentioned that the region with $\theta_2^{\rm ini} \sim 1$ is favored by minimality in the dark sector, as tiny misalignment angels require new physics \cite{Dvali:1995ce,Koutsangelas:2022lte}. Without a discrete symmetry that connects $\Lambda_{\rm conf}$ and $f_{a_2}$ to the limited analogs in our sector, there is simply no need to introduce new physics to achieve tiny misalignment angles. Intriguingly, in the two-form realization, for which the value of $f_{a_2} \sim M_\Prm$ is strongly favored, this minimality-argument results in the dark confinement scale being $\Lambda_{\rm conf} \sim 1\eV$.

\subsection{Isocurvature Perturbations}
\label{sec:Isocurvature}
\vs{-5mm}

When one of the PQ symmetries is broken during (or before) inflation and never restored afterwards, the corresponding axion field is subject to quantum perturbations. Let us write the axion fields as $\theta_i = \VEV{\theta_i} + \delta \theta_i$, where $\VEV{\theta_i}= \theta_i^{\rm ini}$ and $\delta \theta_i$ denotes the quantum perturbations. These fulfill $\VEV{\delta \theta_i}=0$ and have a standard deviation of
	\begin{equation}
		\sigma_{\theta_i}
            \sim
                \sqrt{\VEV{\delta \theta_i^2}}
			\sim
				\frac{H_\Irm}{2 \pi f_{a_i}}
			\; .
	\end{equation}
Furthermore, if none of the axions is responsible for inflation, their fluctuations will not be of adiabatic but of isocurvature-type. Since these lead to a unique imprint in the temperature and polarisation fluctuations of the CMB, they give rise to a constraint on the axion's parameter space.

One could worry that in the presence of several axions, the amount of isocurvature perturbations is significantly enhanced, thus making the existing constraint much more severe. Let us show that this is not necessarily the case, even though the contributions are additive. 

Assuming the perturbations to be normal distributed in the regime of small $\theta_i$, where anharmonic corrections of the potentials can be ignored, the collective amplitude of the axions isocurvature fluctuations is given by \cite{Kobayashi:2013nva}
	\begin{align}\nonumber
		\Delta_{a}(k_0)
			&=
				\frac{\delta \Omega_{\rm DM}}{\Omega_{\rm DM}}
			=
				\frac{\sum_i\Omega_{a_i}}{\Omega_{\rm DM}}
				\frac{\delta \ln \Omega_{a_i}}{\delta \theta_i^{\rm ini}}
				\sigma_{\theta_i} \\
			&=
				\sum_i
                \frac{\Omega_{a_i}}{\Omega_{\rm DM}}	
				\frac{H_\Irm}{\pi \theta_i^{\rm ini} f_{a_i}}
			\; ,	
	\end{align}
where $\theta_i^{\rm ini} \ll \sigma_{\theta_i}$ was used. The latest experimental bound on uncorrelated isocurvature perturbations by Planck is \cite{Aghanim:2018eyx}
	\begin{equation}
		\beta(k_0)
			\equiv
				\frac{ \Delta^2_{a}(k_0)}
				{ \Delta^2_\mathcal{R}(k_0) + \Delta^2_{a}(k_0) }
			<
				 0.038 \quad {\rm at \ 95\% \ CL}
			\; ,
	\end{equation}
where $k_0 = 0.050 {\rm Mpc}^{-1}$. This can be translated to a constrain on $H_\Irm$,
	\begin{equation}
		H_\Irm
			\lesssim
				10^7 \GeV
                \sum_i 
				\frac{\Omega_{\rm DM}}{\Omega_{a_i}}
				\left(
					\frac{f_{a_i}}{10^{12} \GeV}
                \right)
                \left(
					\frac{\theta_i^{\rm ini}}{1}
				\right)
			\; .
    \label{Eq:IsocurvBoundonH}
	\end{equation}
For both models at hand, namely a single dark YM sector and $N$ exact SM copies, the generated isocurvature perturbations are either dominated by our sector or the dark sector(s). 

For the former, if the dark matter is dominated by the axion from the dark sector, we can express the bound on $H_\Irm$ in terms of the dark confinement scale by using the equality of (\ref{Eq:Confinementupperbound}) to eliminate $\theta_2^{\rm ini}$. This results in
    \begin{equation}
		H_\Irm
			\lesssim
				10^7 \GeV
				\left(
					\frac{f_{a_i}}{10^{12} \GeV}
                \right)^{\frac{1}{4}}
                \left(
					\frac{1 \GeV}{\Lambda_{\rm conf}}
				\right)^{\frac{1}{2}}
			\; .
    \label{Eq:IsocurvBoundonH2}
	\end{equation}  
In the case of a two-form axion from such a dark sector, i.e. with $f_{a_2} \sim M_\Prm$ and the minimal possible dark confinement scale $\Lambda_{\rm conf} \sim \eV$, the bound reduces to $H_\Irm \lesssim 10^{13} \GeV$, meaning that such an axion is essentially not plagued by an isocurvature constraint.

For $N$ exact SM copies, when the dark matter is collectively composed by the axions from the copies, i.e. $\sum_i \Omega_{a_i} \sim \Omega_{\rm DM}$, using (\ref{Eq:MisalignmentBoundonN}) to eliminate the initial misalignment angles in (\ref{Eq:IsocurvBoundonH}) yields
	\begin{equation}
		H_\Irm
			\lesssim
				10^7 \GeV
				\left(
					\frac{f_{a_i}}{10^{12} \GeV}
                \right)^{\frac{1}{4}}
                \left(
					\frac{10}{\sqrt{N}}
				\right)
			\; .
	\end{equation}
We observe that for moderate values of $N\sim10^2$ the bound remains relatively unaffected. On the other hand, for values required to solve the Hierarchy problem, i.e. $N \sim 10^{32}$ and $f_a \sim \TeV$, the bound naively tightens to $H_\Irm \lesssim 10^{-10} \GeV$. However, in this case, the dark confinement scale and the axion mass exceed the inflationary Gibbons-Hawking temperature $T_\Irm \sim H_\Irm$, meaning that the dark YM sector becomes strongly coupled during inflation. Consequently, the axion develops a substantial mass so that the requirements for the development of isocurvature perturbations are no longer given. Of course, this requires avoiding the astrophysical bounds on $f_a$ by a mechanism such as clockworking (see discussion at the end of Section~\ref{sec:Exact SM Copies}).

\subsection{$N$-Boson Stars}
\label{sec:N_Axion_Stars}
\vs{-5mm}

Bosonic dark matter is known to be able to form dense clumps, which in the case of the axion are called axion stars. The expression “star” is used to denote an object sustained by hydrostatic equilibrium, regardless of its emission of light. Such objects as well as other variations of compact dark matter structures have been long studied and provide an interesting perspective on the nature of dark matter \cite{TKACHEV1991289} (see \cite{Zhang:2018slz} for a review). 

For an axion arising from a dark YM sector, the physics of axion stars are essentially equal to those of our sector, merely the values of the axion mass and decay constant change. However, the existence of multiple dark sectors may lead to structures that are collectively formed from the axions of the different sectors (see \cite{CompactDarkMatterObjectsViaN} for a realization of this idea with fermions called $N$-MACHOS). These structures would be much lighter and smaller compared to the single-sector case, which can be intuitively understood as follows. When particles from $N$ sectors are placed into a volume, the system is subject to gravity and pressure. Gravity attempts to collapse the system, while pressure from each sector resists the collapse. Eventually, the combined pressure can counteract gravity, resulting in an equilibrium configuration. However, because each particle can only interact with a small portion of the total particles, the overall pressure is much lower than in a single sector case. At the same time, gravity affects all particles equally. As a result, equilibrium can only be achieved with smaller masses and radii than in typical single-sector models. In the following, we argue that this phenomenon can not take place for axions.

The analysis performed in \cite{CompactDarkMatterObjectsViaN} was based on the Tolman-Oppenheimer-Volkoff (TOV) equation, which is a relativistic equation that characterizes the hydrostatic equilibrium structure of a self-gravitating, spherically symmetric object made of a perfect fluid. Unlike perfect fluids, the scalar field of axion stars has a non-trivial self-interaction potential that cannot be described by a simple equation of state. As a result, the equilibrium structure of axion stars cannot be determined solely by the TOV equation. Instead, it requires the solution of the Klein-Gordon equation that describes the scalar field, which must be solved simultaneously with the Einstein field equations. 

Although a comprehensive analysis would certainly be of interest, here, we limit ourselves to so called dilute axion stars. To identify such configurations, we follow the standard approach (see for instance \cite{Zhang:2018slz}) and employ a set of approximations to simplify the relevant equations while still maintaining a high level of accuracy. We aim to maintain a maximal level of generality in our calculations to allow for the application to scalars with various potentials.

To begin, we can utilize non-relativistic effective field theory since the axions produced through misalignment are non-relativistic. In this framework, it is convenient to introduce the complex field $\psi_i(x)$ for each axion through 
    \begin{equation}
        a_i(x)
            =
                \frac{1}{\sqrt{2 m_{a_i}}}
                \left[
                    \psi_i(x) \erm^{i m_{a_i} t}
                    + \psi_i^*(x) \erm^{-i m_{a_i} t}
                \right]
            \; .
    \end{equation}
By taking the non-relativistic limit of each Klein-Gordon equation and expressing the axion field in terms of $\psi_i$, we obtain the time-dependent Gross-Pitaevskii equation. To furthermore separate the time dependence, we make the standard separation ansatz $\psi_i(\Vec{x},t) = A_i(\Vec{x}) \erm^{-i\omega_i t}$. This results in the time-independent Groß-Pitaevskii equation,
    \begin{equation}
        m_{a_i}
        \left(
            \Phi + V_i'(|\psi_i|^2)
        \right) 
        - \frac{1}{2m_{a_i}}
        \frac{\Delta \psi_i}{\psi_i}
            =
                E_i
            \; .
    \end{equation}
Next, the cosmic axion condensate has a relatively low mean mass density, so we can use Newtonian gravity described by the Poisson equation,
    \begin{equation}
        \Delta \Phi 
            =
                \frac{\sum_i m_{a_i} |\psi_i|^2}{M_\Prm^2}
            \; .
    \end{equation}
Crucially, the sum appears because gravity couples to all the axions. Within these approximations, the Groß-Pitaevskii-Poisson equations provide a simplified set of equations that can accurately describe dilute axion stars. We prefer to rewrite this system in the form of hydrodynamic equations. To do so, we first take the Laplacian of the Gross-Pitaevskii equations and plug in the Poisson equation. Secondly, we perform the Madelung transformation to express everything in terms of the (pseudo) density $\rho_i = m_{a_i} |\psi_i|^2$ and the pressure, which is related to the potential via $\nabla V_i'(\rho_i) = \nabla p_i(\rho_i) / \rho_i$. This results in the fundamental equation of hydrostatic equilibrium with quantum effects,
    \begin{equation}
        - \nabla \cdot
        \left( 
            \frac{ \nabla p_i}{\rho_i}
        \right)
        + \frac{1}{2m_{a_i}^2} \Delta 
        \left( 
            \frac{\Delta \sqrt{\rho_i}}{\sqrt{\rho_i}}
        \right)
            =
                \frac{\rho_{\rm tot}}{M_\Prm^2}
            \; .
    \label{Eq:HydroEquil}
    \end{equation}
The first term accounts for the hydrodynamic pressure, which can either be attractive or repulsive, while the second term characterizes the repulsion from quantum pressure due to the Heisenberg uncertainty principle. On the right hand side the total density, $\rho_{\rm tot}=\sum_i \rho_i$, appears because gravity couples to all axions.

The last approximation regards the (pseudo) density $\rho$, which we take to be small compared to the cosmic condensate density $m_{a_i}^2 f_{a_i}^2$. In this regime the effective potential $V_i(\rho_i)$ is dominated by the leading term in its power series. For the instantonic potential from (\ref{Eq:PotDIGA}), the leading order interaction is the quartic term of $a_i$, so in terms of $\rho_i$ the potential can be approximated as
    \begin{equation}
        V_i(\rho_i)
            \sim 
                - \frac{1}{m_{a_i}^2 f_{a_i}^2} \rho_i^2 
            \; .
    \end{equation}
This is equivalent to the negative polytropic pressure
    \begin{equation}
        p_i 
            \sim
                - \frac{1}{m_{a_i}^2 f_{a_i}^2} \rho_i^2
            \equiv  
                K_i \, \rho_i^{\gamma_i}
            \; ,
    \end{equation}
with polytropic constant $K_i \sim - 1/(m_{a_i}^2 f_{a_i}^2)$ and polytropic index $\gamma_i=2$. With such an equation of state, (\ref{Eq:HydroEquil}) becomes
    \begin{equation}
        - \nabla 
        \left( 
            \gamma_i K_i \rho_i^{\gamma_i-2} \nabla \rho_i
        \right)
        + \frac{1}{2m_{a_i}^2} \Delta 
        \left( 
            \frac{\Delta \sqrt{\rho_i}}{\sqrt{\rho_i}}
        \right)
            =
                \frac{\rho_{\rm tot}}{M_\Prm^2}
            \; .
    \label{Eq:HydroEquilPoly}
    \end{equation}

For the sake of illustration, let us assume that all densities are equal, 
    \begin{equation}
        \rho_i 
            = 
                \rho_{\rm tot}/N
            \; ,
    \end{equation}
as collective effects are most pronounced when all sectors have equal densities \cite{CompactDarkMatterObjectsViaN}. In this way, we can express (\ref{Eq:HydroEquilPoly}) in terms of the total density, 
    \begin{equation}
        - \nabla 
        \left( 
            \frac{\gamma_i K_i \rho_{\rm tot}^{\gamma_i-2}}{N^{\gamma_i-1}} \nabla \rho_{\rm tot}
        \right)
        + \frac{1}{2m_{a_i}^2} \Delta 
        \left( 
            \frac{\Delta \sqrt{\rho_{\rm tot}}}{\sqrt{\rho_{\rm tot}}}
        \right)
            =
                \frac{\rho_{\rm tot}}{M_\Prm^2}
            \; .
    \end{equation}
We observe that in the presence of $N$ axions the pressure from short-range interactions is suppressed by $N^{\gamma_i-1}$ while the quantum pressure is unaffected. 

The influence of $N$ on the solution can be understood by considering the following limiting cases. 
    \begin{itemize}
        \item Neglecting the hydrostatic pressure, the equilibrium is between the gravitational attraction and the repulsion by the quantum pressure, which are both independent of $N$. Thus, the final configuration is unchanged from the single axion case.
        \item Neglecting the quantum pressure, the equilibrium is between the gravitational attraction and the hydrostatic pressure (which must be repulsive), which is now strongly suppressed by $N^{\gamma_i-1}$. From here on the analysis is exactly the same as in \cite{CompactDarkMatterObjectsViaN}, so that the radius and mass of the final configuration would be suppressed by $1/\sqrt{N}$.
        \item Neglecting the gravitational attraction, the equilibrium is between the hydrostatic pressure (which must be attractive) and the repulsion by quantum pressure. This would again result in an altered mass spectrum, however, such equilibria are always unstable \cite{Chavanis_2011}.
    \end{itemize}
We can see that stable collective configurations with a suppressed mass spectrum can only exist for repulsive interactions, which is not the case for the axion. Therefore, dilute axion stars would consist of several axions but would not differ in mass or radius compared to the single sector case. However, we want to stress that they would exist for scalars with a repulsive interaction (see for instance \cite{Berezhiani:2021rjs, Berezhiani:2022buv})

\section{Kinetic Mixing between Axions}
\label{sec:Kinetic_Mixing}
\vs{-5mm}

\subsection{Kinetic Mixing as Intersector Interaction}
\label{sec:Kinetic_Mixing}
\vs{-5mm}

In this section, we discuss non-gravitational intersector interactions by restoring $\mathcal{L}_{\rm mix}$ in (\ref{Eq:GeneralLagrangian}). For now, we again put the focus on $N$ exact SM copies. Without the additional axions, the possible renormalizable interactions of this model, which are compatible with gauge-, Lorentz-, and the underlying discrete symmetry, are photon kinetic mixing, a Higgs portal coupling, and neutrino mass mixing. At the non-renormalizable level, additional interactions such as neutron oscillations become possible as well. The effects of these interactions have been vastly discussed in the literature (see \cite{Foot:2014mia} for a review). As our focus is not on these interactions, we will not elaborate further on their influence. Instead, we focus on axion kinetic mixing, which in the model under consideration is described by  
    \begin{equation}
        \mathcal{L}_{\rm mix}
            =
                \epsilon \sum_{i \neq j} \partial_\mu a^i \partial^\mu a^j
            \; ,
    \label{Eq:LagrangianKinMix}
    \end{equation}
where $\epsilon$ parametrizes the kinetic mixing strength. An important implication of kinetic mixing between our visible and hidden sectors is that it can result in thermal equilibrium between the sectors. In order to prevent inconsistencies with nucleosynthesis or an excessive amount of dark matter particles, the thermal equilibrium must not occur prior to BBN, usually leading to a constraint on the kinetic mixing strength. The axions that are produced by misalignment are non-thermal. Thus, they can not transfer heat between the sectors via axion-axion interactions to thermalize any sectors. However, this is not true for other types of interactions, potentially leading to a constraint on $\epsilon$ in order to avoid reheating of the dark sectors prior to nucleosynthesis. Such an analysis, which requires further assumptions, goes beyond the scope of this paper and we simply regard $\epsilon$ as a (phenomenologically) free parameter in the following. 

\begin{figure}[t]
        \centering
        \includegraphics[scale= 0.4]{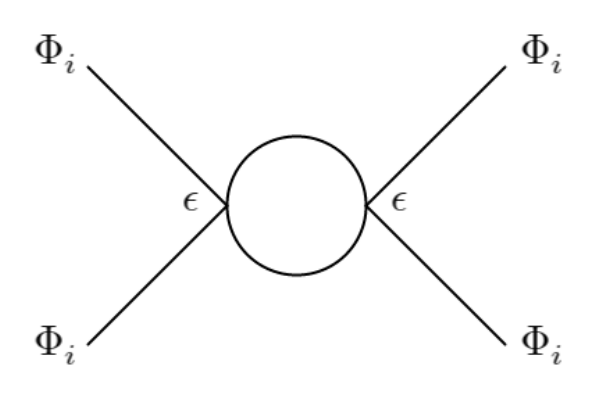}
        \caption{Loop induced by the operators in (\ref{Eq:LagranianKinMixUV}). Each vertex contributes the effective coupling $\Delta p^2/M^2$, but for maximum momentum transfers of order $f_{a}$ the effective coupling becomes $\epsilon$.}
        \label{Fig:Loop}
    \end{figure}

The way in which axion kinetic mixing can emerge in the low energy effective theory depends on the specific implementation of the axion. In the standard implementation of the axion via the KSVZ model a singlet Higgs is added to make the axion invisible. Denoting this singlet in each sector as $\Phi_i$ allows for the following dimension six Higgs portal couplings between the sectors
    \begin{equation}
        \mathcal{L}_{\rm mix}^{\rm UV}
            =
                \frac{1}{M^2} 
                \sum_{i \neq j}
                (\Phi_i^\dagger \partial_{\mu}\Phi_i)
                (\Phi_j \partial^{\mu} \Phi_j^\dagger)
                + {\rm h.c.}
            \; ,
    \label{Eq:LagranianKinMixUV}
    \end{equation}
where $M$ is some cut-off scale. The singlets acquire the VEVs $f_{a_i} \equiv f_a$ via a proper scalar potential that spontaneously breaks the U(1)$_\PQ$ symmetries. This results in the kinetic mixing described in (\ref{Eq:LagrangianKinMix}) with
    \begin{equation}
        \epsilon 
            \equiv 
                \frac{f_a^2}{2 M^2}
            \; .
    \end{equation}
A diagram coming from such a operator is shown in Figure \ref{Fig:Loop}. Similar operators that result in axion kinetic mixing are also possible in DFSZ-type models, where extra Higgs doublets are required in addition to the singlet.

While we neglect the phenomenological constraint on $\epsilon$ arising from the necessary avoidance of the dark sector thermalization before BBN, it is still subject to consistency limitations due to unitarity. The operators in (\ref{Eq:LagranianKinMixUV})
 already at tree-level put a unitarity constraint:
    \begin{equation}
        \epsilon 
            \lesssim 
                \frac{1}{\sqrt{N}}
            \; .
    \label{Eq:UnitarityBound2}
    \end{equation}

However, the bound on $\epsilon$ is much more severe. Since it describes an intersector interaction coupling at momentum transfer $q \sim f_a$ among all species, it has to fulfill the bound \eqref{speciescouping} of such a coupling imposed by the entropy of species \cite{Dvali:2020wqi}. The bound \eqref{speciescouping} gives: 
    \begin{equation}
        \epsilon 
            \lesssim 
                \frac{1}{N}
            \; .
    \label{Eq:UnitarityBound}
    \end{equation}

In the alternative two-form implementation of the axion, axion kinetic mixing appears via non-diagonal mass terms of three-form action,
    \begin{equation}
        \mathcal{L}_{\rm mix}
            \sim
                \sum_{i \neq j}
                \left(
                    C^i_{\mu\nu\rho} - \partial_{[\mu} B^i_{\nu\rho]}
                \right)
                \left(
                    C^j_{\mu\nu\rho} - \partial_{[\mu} B^j_{\nu\rho]})
                \right)
            \; .
    \end{equation} 
Mixing of this kind is not forbidden by gauge symmetries and can originate from virtual black hole exchange \cite{Dvali:2007iv}, but there may be other origins. The emergence of such operators is for $i \neq j$ only possible if they are suppressed by powers of $M_\Prm^{-1}$. The reason for this suppression is that micro black holes cannot be universally coupled, which necessitates gravitational suppression of inter-sector transitions at the fundamental level. 

In the remainder of this section, we will study the consequences of the axion kinetic mixings in the pseudo-scalar formulation. We will stay agnostic of their origin, but use the unitarity bound described by (\ref{Eq:UnitarityBound}).

\subsection{Modification of Axion Physics for SM Copies}
\label{sec:Axion-Photon Coupling}
\vs{-5mm}

The presence of the extra term in the Lagrangian necessitates a change of basis to express it in a canonical form. As a result, there is a mismatch between the ``sector basis", which corresponds to the labels of the species, and the ``canonical kinetic basis", in which the propagator is in canonical form. So the task is to find the relation between these two bases.

Before discussing this change of basis, it is worth noting that a similar situation arises in the SM's neutrino sector. There, the interaction terms are diagonal in the flavor basis. But in this basis the mass matrix of neutrinos is off-diagonal. Therefore, there is a mismatch between the flavor and the mass basis, which leads to phenomenons like neutrino oscillations. In our case, the mismatch appears between the canonical kinetic basis and the sector basis.

The first step is to express the kinetic part in (\ref{Eq:LagrangianKinMix}) as
\begin{equation}
    \mathcal{L} 
        \supset 
            \begin{pmatrix} 
                \partial_{\mu}a_1 \\ 
                \vdots \\ 
                \vdots \\ 
                \partial_{\mu} a_N
             \end{pmatrix}^T 
            \begin{pmatrix} 
                1&& \epsilon&& \dots && \epsilon \\ 
                \epsilon&& 1&& \ddots && \vdots \\ 
                \vdots&& \ddots &&\ddots&&\epsilon\\ 
                \epsilon&&  \dots&& \epsilon &&1
            \end{pmatrix} 
            \begin{pmatrix} 
                \partial_{\mu}a_1 \\ 
                \vdots \\ 
                \vdots \\ 
                \partial_{\mu} a_N
            \end{pmatrix} 
    \; .
     \label{rewritten}
\end{equation}
This matrix appears due to the permutation symmetry among the copies also in other cases, which have been studied in \cite{Dvali:2009ne}. We can rewrite the matrix, which we shall call $K$ from now on, in the following way,
    \begin{equation} 
        K
            =
                \begin{pmatrix}
                    1-\epsilon &0&\dots&0\\
                     0&1-\epsilon&\ddots&\vdots\\
                     \vdots&\ddots&\ddots&0 \\
                     0&\dots& 0&1- \epsilon
                \end{pmatrix}   
                + \epsilon 
               \begin{pmatrix}
                    1 & \dots&1  \\
                    \vdots&\ddots&\vdots\\
                    1 &\dots& 1  
                \end{pmatrix}  
            \; .
    \end{equation}
The problem reduces then to the diagonalization of a matrix of just ones. The transformation matrix $S$ that diagonalizes $K$ is
    \begin{equation}
            S 
                =
                    \begin{pmatrix}
                        1 & 1 & 1 & \dots & 1\\
                        1 & -1 & 0 & \dots & 0\\
                        1 & 0 & -1 & \ddots & \vdots\\
                        \vdots & \vdots & \ddots & \ddots & \vdots\\
                        1 & 0 & \cdots & 0 & -1
                    \end{pmatrix}
                \; .
        \end{equation}
The first row gives rise to the eigenstate of the form,
        \begin{equation}
            \Tilde{a}_L 
                = 
                    \frac{1}{\sqrt{N}} a_1 
                    + \sqrt{\frac{N-1}{N}} a_h
                \; ,
                \label{Eq:LightState}
        \end{equation}
where we have introduced the notation
    \begin{equation}
       a_h 
            = 
                \frac{1}{\sqrt{N-1}}\sum_{i=2} a_i
            \; ,
    \end{equation}
due to later convenience. The eigenstate $\Tilde{a}_H$ corresponds to the eigenvalue of $1+(N-1)\epsilon$. Because the matrix $S$ is not a unitary matrix we still have to find a convenient basis.

From all other rows, we see that we have $N-1$ degenerate eigenstates $v_i$ of the eigenvalue $1 - \epsilon$. These $v_i$´s are the columns of $S$. Due to this degeneracy, we can reduce the $N \times N$ problem to a $2 \times 2$ problem by defining a superposition of these degenerate eigenstates. This looks like
    \begin{equation}
       \Tilde{a}_H =\frac{1}{N-1} \sum_i v_i = a_1 - \frac{1}{\sqrt{N-1}} a_h
       \; ,
    \end{equation}
and again has the eigenvalue $1 - \epsilon$. After normalization this becomes
    \begin{equation}
        \Tilde{a}_H = \sqrt{\frac{N-1}{N}}a_1 - \frac{1}{\sqrt{N}}a_h
        \; .
     \label{a_H}
    \end{equation}
We see that $\tilde{a}_H$ is a collective expression made out of all former $a_i$´s. Using expressions (\ref{a_H}) and (\ref{Eq:LightState}), we find $a_1$ expressed in the canonical kinetic basis
    \begin{equation}
        a_1 = \sqrt{\frac{N-1}{N}} \Tilde{a}_H + \frac{1}{\sqrt{N}}\Tilde{a}_L
        \; .
        \label{speciesstate}
    \end{equation}
Note that the choice of what we label as species one is totally arbitrary and therefore holds for every $i = 1,...,N$. This results from having the same parameter $\epsilon$ among all sectors. The $2 \times 2$ matrix connecting $a_1$ and $a_h$ with $\tilde{a}_H$ and $\tilde{a}_L$ is now a unitary matrix.

Expressing (\ref{rewritten}) in terms of $\tilde{a}_L$ and $\tilde{a}_H$, there is no more mixing but each term is multiplied by the corresponding eigenvalue. In order to have canonical kinetic terms, the states $\tilde{a}_L$ and $\tilde{a}_H$ need to be redefined by
    \begin{equation}
        \Tilde{a}_H \rightarrow \frac{1}{\sqrt{1-\epsilon}} \Tilde{a}_H 
        \; ,
        \quad 
        \Tilde{a}_L \rightarrow \frac{1}{\sqrt{1+(N-1)\epsilon}} \Tilde{a}_L
        \; .
        \label{redefinition}
    \end{equation}

One effect of this redefinition is that the mass part in the Lagrangian is changed to 
    \begin{equation}
            \mathcal{L}_{\rm mass} \sim          
             \begin{pmatrix} 
                \Tilde{a}_H \\ 
                \Tilde{a}_L
             \end{pmatrix}^T 
            \begin{pmatrix} 
                \frac{m_a^2}{1-\epsilon}&& 0\\ 
                0&& \frac{m_a^2}{1+(N-1)\epsilon} 
            \end{pmatrix} 
            \begin{pmatrix} 
                \Tilde{a}_H \\ 
                \Tilde{a}_L
             \end{pmatrix}
    \; .     
    \label{Eq:MassLagrangianSMcop}
    \end{equation}
with $m_a$ being the mass induced by the PQ-mechanism. One sees that kinetic mixing leads to a splitting of the masses of the axions. $N-1$ axions are degenerated and one light axion whose mass is suppressed by the number of copies. In other words, kinetic mixing of many copies of the axion leads to two different detectable axion states which have different masses. The relation between these masses is 
\begin{equation}
    \frac{m_L}{m_H} 
        = 
            \sqrt{
                \frac{1-\epsilon}{1 + (N-1)\epsilon} 
            }
        \sim 
            \frac{1}{\sqrt{2}}
        \; ,
    \label{ratioofmasses}
\end{equation}
where for the approximation we assumed natural values of $\epsilon$ with respect to the unitarity bound, i.e. $\epsilon \sim 1/N$, and $N \gg 1$.

    \begin{figure*}[t]
        \includegraphics[scale= 0.45]{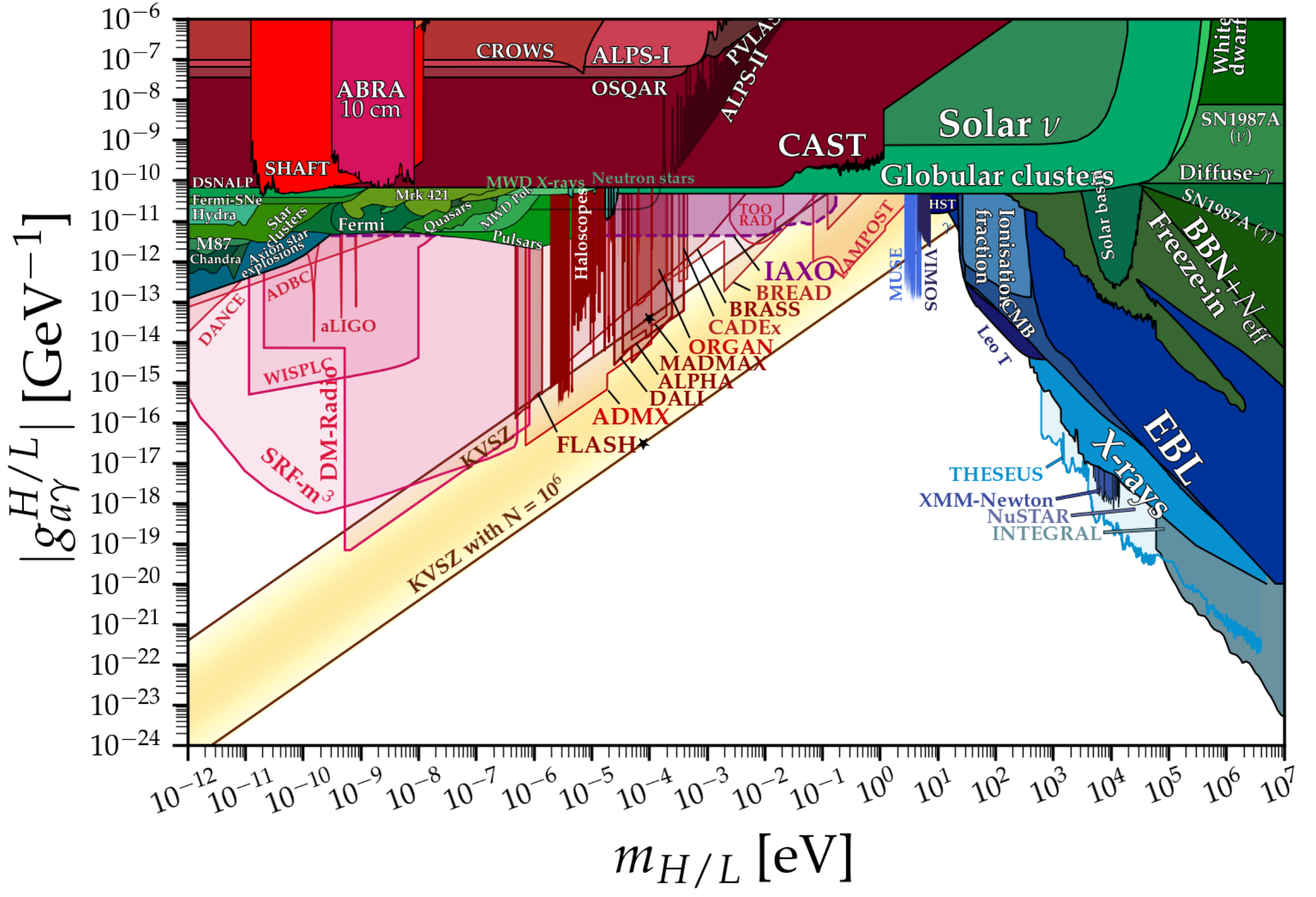}
        \caption{Viable axion-photon couplings for the light axion state (yellow band), which arises from equal kinetic mixing of axions from $N$ sectors. Current bounds and predicted sensitivities of future experiments are depicted in different colors. The parameter space for $N \lesssim 10$ will be probed by future experiments. The yellow band is not to be confused with the bands from \cite{DFSZ}, which include the set of DFSZ-type or KSVZ-type models. The two stars represent a situation when future experiments measure one axion (the upper star) and where we expect the second axion (the lower star). This plot was created with the help of the software \cite{AxionLimits}.}
        \label{AxionPhoton_with_Projections}
    \end{figure*}

With the relation between the species basis and the canonical kinetic basis, we can turn to the phenomenological implications of this model. Due to the Goldstone nature of the axion, its lowest order couplings to fermions and gauge bosons have the generic form 
    \begin{equation}
        g_{ao} a_1 \mathcal{O} \; ,
    \end{equation}
where the whole UV dependency is encoded in $g_{ao}$. After the diagonalization and field redefinition of (\ref{redefinition}), the coupling becomes,
    \begin{equation}
       g_{ao} \left(
           \sqrt{\frac{N-1}{N}}\frac{1}{\sqrt{1-\epsilon}}\Tilde{a}_H + \frac{1}{\sqrt{N}}\frac{1}{\sqrt{1+(N-1)\epsilon}} 
           \Tilde{a}_L
       \right)
       \mathcal{O}
       \; .
    \label{coupling}
    \end{equation}
We observe that by having $N$ kinetically mixed axions, our sector couples to $N$ axions instead of one. These $N$ axions come in two categories: $N-1$ axions ($\tilde{a}_H$) that behave exactly the same, and one special axion ($\tilde{a}_L$). The single axion coupling $g_{ao}$ gets modified by one of the following factors,
    \begin{equation}
        f_H(N,\epsilon)
            \equiv
                \sqrt{\frac{N-1}{N}}\frac{1}{\sqrt{1-\epsilon}}
            \sim 
                1
            \; , \\
            \label{couplingheavyaxion}
    \end{equation}
    \begin{equation}
        f_L(N,\epsilon)
            \equiv
                \frac{1}{\sqrt{N}}\frac{1}{\sqrt{1+(N-1)\epsilon}}
            \sim
                \frac{1}{\sqrt{2N}}
            \; ,
                \label{couplinglightaxion}
    \end{equation}
where for the approximation we again used the natural value of $\epsilon$ and $N \gg 1$. We observe that the coupling of $\tilde{a}_H$ remains roughly the same as in the single axion case, while $\tilde{a}_L$ shows a suppression. Overall, we can say that on top of the above mentioned mass splitting into two physical states, their couplings differ by the factors $f_H$ and $f_L$.

For the axion-photon coupling, this modification results in 
    \begin{align}\nonumber
        g_{a\gamma}^{H/L}
            &= 
                \frac{\alpha}{2\pi f_a}
                \left(
                    \frac{\mathcal{E}}{\mathcal{N}}
                    - 1.92
                \right)
                f_{H/L}(N, \epsilon) \\
            &= 
                \frac{\alpha}{2\pi f_a}
                C_{a\gamma}
                f_{H/L}(N, \epsilon)
            \; ,
    \label{axionphotonequation}
    \end{align}
where $\mathcal{E}$ and $\mathcal{N}$ are the electromagnetic and QCD anomaly coefficients, respectively, and $\alpha$ is the electromagnetic fine-structure constant. In the second line, we defined the model dependent factor $C_{a\gamma}$ for compactness. 
In order to bring this to a more useful form, we express $f_a$ in terms of the physical mass states from (\ref{Eq:MassLagrangianSMcop}), i.e. 
    \begin{equation}
        m_a = \frac{\sqrt{\Lambda_{\rm QCD}^3 m_u}}{f_a} = \begin{cases}
                  \sqrt{1-\epsilon} \ m_H
                    &: {\rm for} \ \Tilde{a}_H
                
                \; , 
                \\
                \sqrt{1+(N-1)\epsilon} \ m_L
                \; &: {\rm for} \ \Tilde{a}_L.
                \end{cases}
    \end{equation}
After plugging in (\ref{couplingheavyaxion}) and (\ref{couplinglightaxion}), this results in 
    \begin{align}
        \label{Eq:HeavyAPCoupling}
        g_{a\gamma}^{H}
            &= 
                \frac{\alpha}{2\pi}
                \frac{ C_{a\gamma}}{ \sqrt{\Lambda_{\rm QCD}^3 m_u}}
                \sqrt{ \frac{N-1}{N} } \ m_H \\
        \label{Eq:LightAPCoupling}
        g_{a\gamma}^{L}
            &= 
                \frac{\alpha}{2\pi}
                \frac{ C_{a\gamma}}{ \sqrt{\Lambda_{\rm QCD}^3 m_u}}
                \sqrt{ \frac{1}{N} } \ m_L
            \; .
    \end{align}

For concreteness, let us investigate the axion-photon couplings in the case of KSVZ axions. In Figure~\ref{AxionPhoton_with_Projections} we show the possible couplings for $N < 10^{6}$. The included projected sensitivities of future experiments is ranging into the predicted band but most of the parameter space will not be covered in the near future. Nevertheless, one has to keep in mind that the heavy axion, which couples similarly to a scenario with just one axion (maximal enhancement is a factor of $\sim 1.3$), has a realistic chance of being discovered in the near future. In the framework of multiple axions, the next step after the discovery of the ordinary axion would be to search for a second light weakly coupled state. Even though one could think that the task is quite difficult because the available parameter space is large, one has to keep in mind that there is a clear functional dependence among the couplings and the masses between the heavy and the light state. So imagine a situation that an axion has been discovered and therefore $m_H$ and $g_{a \gamma}^H$ would be known. In the expression for the coupling of the light axion (\ref{Eq:LightAPCoupling}), we can replace $m_L$ by $m_H$ using (\ref{ratioofmasses}). We can then eliminate $N$ by plugging in the expression for the heavy axion (\ref{Eq:HeavyAPCoupling}), yielding an expression for $g_{a \gamma}^L$ in terms of experimentally known parameters,
    \begin{align}
        g_{a \gamma}^L 
            \sim 
                \frac{\alpha C_{a\gamma} m_H}{2 \pi \sqrt{\Lambda_{\rm QCD}^3 m_u} }
                    \sqrt{1- \left( 
                    \frac{2 \pi g_{a \gamma}^H  \sqrt{\Lambda_{\rm QCD}^3 m_u}}
                    {\alpha C_{a\gamma} m_H} \right)^2}
            \; ,
    \end{align}  
This means that after measuring the coupling and the mass of the first axion, the properties of the second are uniquely determined. 

\subsection{Modification of Axion Physics for one YM Sector}

Another interesting case for kinetic mixing is the scenario of a single YM sector with PQ scale $f_{a_2}$ and confinements scale $\Lambda_{\rm conf}$, as discussed in Section~\ref{sec:Pure_YM}. The procedure of analyzing the kinetic mixing remains unchanged and $N = 2$ in this case. 
    
\begin{figure}[t]
        \centering
        \includegraphics[width=0.483\textwidth]{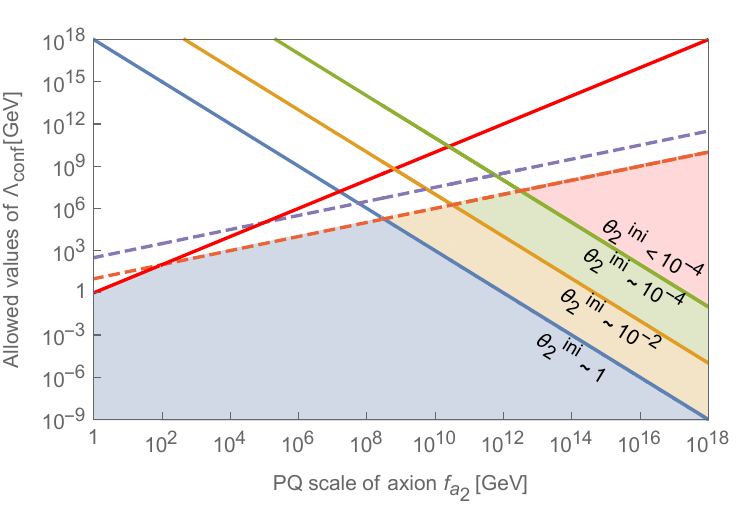}
        \caption{Same as Figure.~\ref{fig:Lambdai} but with axion kinetic mixing. The mixing results in both axion states being able to decay into our photons. Requiring the stability of the heavy axion state results in the viable region to lie below the dashed orange line ($f_a \sim 10^{12} \GeV$) or below the dashed purple line ($f_a \sim M_\Prm$).
        }
    \label{Fig:AxionPhoton}
    \end{figure}

The coupling to our photons after the step of choosing the proper kinetic basis reads as
    \begin{equation}
        g_{a\gamma}^{L/H}
            = 
                \frac{\alpha}{2\pi f_a}
                C_{a\gamma}
                \frac{1}{\sqrt{2}}
                \frac{1}{\sqrt{1 \pm \epsilon}}
            \; .
    \label{Eq:AxionPhotonCoup}
    \end{equation}
Since $\Lambda_{\rm conf}$ and $f_{a_2}$ differ from the analog quantities in our sectors, the axion masses differ as well. This leads to a non-diagonal mass matrix,
\begin{align} \nonumber
    \mathcal{L}_{\rm mass} 
        &\sim m_{a_1}^2 a_1 a_1 +m_{a_2}^2 a_2 a_2  \\
        &\sim 
            \begin{pmatrix} 
                \Tilde{a}_H \\ 
                \Tilde{a}_L
             \end{pmatrix}^T \begin{pmatrix} 
                \frac{m_{a_1}^2 + m_{a_2}^2}{1-\epsilon}&& \frac{m_{a_1}^2-m_{a_2}^2}{\sqrt{1-\epsilon}\sqrt{1+\epsilon}}\\ 
                \frac{m_{a_1}^2-m_{a_2}^2}{\sqrt{1-\epsilon}\sqrt{1+\epsilon}}&& \frac{m_{a_1}^2 + m_{a_2}^2}{1+\epsilon}
            \end{pmatrix} 
            \begin{pmatrix} 
                \Tilde{a}_H \\ 
                \Tilde{a}_L
             \end{pmatrix}.
\end{align}
As usual, this mass-term must be diagonalized by an orthogonal matrix with the mixing angle
\begin{equation}
    \theta 
        = 
            \frac{1}{2}\arctan\left( 
                \frac{m_{a_1}^2-m_{a_2}^2}{m_{a_1}^2+m_{a_2}^2}
                \frac{\sqrt{1-\epsilon^2}}{\epsilon}
            \right)
        \; ,
\end{equation}
leading to the eigenvalues,
    \begin{align}
    m_{H/L} 
        =
            \frac{
                m_{a_1}^2 + m_{a_2}^2 \pm (m_{a_1}^2 - m_{a_2}^2)
                \sqrt{
                    1 + \frac{4\epsilon^2}{\left(\frac{m_{a_1}}{m_{a_2}}-\frac{m_{a_2}}{m_{a_1}}\right)^2}
                }
            }{1-\epsilon^2}
    \; .
    \label{Eq:MassEigenstatesFinal}
    \end{align}
Expressing $a_H$ and $a_L$ in terms of the final eigenstates $A_H$ and $A_L$, the axion-photon coupling from (\ref{Eq:AxionPhotonCoup}) becomes
    \begin{align} \nonumber
        g_{A\gamma}^{H/L}
            &= 
                \frac{\alpha}{4\pi f_{a_1}}
                C_{a\gamma}
                \left(
                    \frac{\cos{\theta}}{\sqrt{1+\epsilon}}
                    \pm \frac{\sin{\theta}}{\sqrt{1-\epsilon}}
                \right) \\
            &\equiv
                \frac{\alpha}{4\pi f_{a_1}}
                C_{a\gamma}
                \kappa_{H/L}
            \; .        
    \end{align}
Using this coupling and the masses, given in (\ref{Eq:MassEigenstatesFinal}), the lifetime of the axions is given by
    \begin{align}
        \tau(A_{H/L} \rightarrow 2\gamma) 
            =
                \frac{2^6 \pi}{\left(g_{A\gamma}^{H/L}\right)^2 m_{H/L}}
            = 
                \frac{2^{10} \pi^3}{\alpha^2 
                C_{a\gamma}^2}
                \frac{f_{a_1}^2}{\kappa_{H/L}^2 m_{H/L}^3} 
            \; .
    \end{align}
For the axion to constitute a part of the dark matter, $\tau(A_{H/L} \rightarrow 2\gamma)$ must be larger than the age of the universe. This requirement translates to the viable region shown in Figure \ref{Fig:AxionPhoton}. While we chose $\epsilon\sim1/2$ in this figure, the viable region is essentially independent of $\epsilon$.

The interpretation of this viable region is vastly unchanged from the discussion in Section~\ref{sec:Pure_YM}. With kinetic mixing an additional upper bound of the dark confinement scale emerges. Consequently, if the dark sector axion makes up the dark matter, the confinement scale of the dark YM sector must be in the range
    \begin{equation}
        1 \eV \lesssim \Lambda_{\rm conf} \lesssim 10^{12} \GeV
        \; .
    \end{equation}

\section{Conclusions}
\label{sec:Summary_and_Outlook}
\vs{-5mm}

In this paper, we studied phenomenological consequences of adding axions to dark YM sectors. These axions are required from quantum gravitational arguments that promote the strong CP problem to a consistency problem. As such, an axion becomes a necessary component not only in QCD but in every YM group \cite{Dvali:2018dce, dvali2022strong}. We focused on two models, namely $N$ exact copies of the SM \cite{Dvali:2007hz, Dvali:2007wp} and a single pure YM sector.

We first required that the total axion density including the axions from all sectors must not exceed the observed dark matter density. For $N$ exact SM copies, using the misalignment mechanism in all sectors results in a limited range of viable $N$ for a given PQ scale. Since the misalignment mechanism is independent of the thermalization of the dark sectors, this applies even for very large values of $N$ when the sectors are very dilute. For a single pure YM sector, we find a relation between the dark confinement scale and the PQ scale. In particular, requiring the dark sector axion to make up the dark matter, the PQ scale does not exceed the Planck scale when the dark confinement scale fulfills  $\Lambda_{\rm conf} \gtrsim 1 \eV$ in the scenario with maximal misalignment. This minimal value is pushed higher for smaller initial misalignment angles. 

Furthermore, we showed that the contribution of additional axions to the isocurvature fluctuations does not necessarily tighten the bound on $H_\Irm$. In fact, both models exhibit regions in the parameter space where the bound is essentially avoided. For $N$ exact SM copies the viable parameter space requires either a moderate number of $N\sim 10^2$ copies or a mechanism that allows for smaller PQ scales.

In models with several dark sectors, it is possible that particles from different sectors collectively form structures with a mass spectrum suppressed by $1/\sqrt{N}$  \cite{Dvali:2019ewm}. While this phenomenon was originally studied with fermions, we showed that it also applies to bosons with a repulsive self-interaction in the regime of dilute boson stars. Since axions have attractive self-interactions, this phenomenon cannot take place. Therefore, the presence of multiple axions does not modify the mass spectrum of such axion stars.

Lastly, we studied non-gravitational communication between the dark sectors from axion kinetic mixing. We showed that a mass splitting into $N-1$ degenerate states and one light state emerges. Interestingly, the latter is lighter by a factor of $1/\sqrt{2}$ and has couplings weaker by a factor of $1/\sqrt{N}$ with respect to the heavier axion states. For $N$ exact SM copies, this allows us to completely predict the light axion once the heavier one is discovered. For a single YM sector, the axion masses are different, so that the axion from the dark sector could decay into photons. If this axion is supposed to make up the dark matter, an upper bound on the dark confinement scale appears by requiring the axion's stability, i.e. $\Lambda_{\rm conf} \lesssim 10^{12}\GeV$. This bound is for most parts of the parameter space stronger than the bound $\Lambda_{\rm conf} \lesssim f_{a_2}$ arising from perturbative unitarity.

The present paper provides certain insights into the potential consequences of multiple axions. As mentioned earlier, the situation with multiple axions closely resembles the axiverse proposed by string theory \cite{Arvanitaki:2009fg}. However, in our case, the axions arise from quantum gravitational consistency in theories with numerous hidden YM groups. Given that hidden YM groups and multiple axions are predictions of string theory, exploring this connection further is a valuable pursuit. This connection can in principle also be applied to dark energy. If the dark energy is realized by quintessence \cite{Ratra:1987rm, Wetterich:1987fm}, it would suggest the existence of a non-trivial vacuum structure that needs to be eliminated by this scalar field. All in all, it seems that axions could play a larger role in nature than initially anticipated.

\acknowledgments
\vs{-5mm}

We are grateful to Gia Dvali and Georg Raffelt for very useful discussions and helpful comments on the manuscript.
This work was partly supported by the Sonderforschungsbereich SFB1258.	
	
\setlength{\bibsep}{5pt}
\setstretch{1}
\bibliographystyle{utphys}
\bibliography{refs}

\providecommand{\href}[2]{#2}\begingroup\raggedright\begin{thebibliography}{10}

\bibitem{ThetaBound}
J.~M. Pendlebury {\em et~al.}, ``{Revised experimental upper limit on the
  electric dipole moment of the neutron},''
  \href{http://dx.doi.org/10.1103/PhysRevD.92.092003}{{\em Phys. Rev. D}
  {\bfseries 92} no.~9, (2015) 092003},
  \href{http://arxiv.org/abs/1509.04411}{{\ttfamily arXiv:1509.04411
  [hep-ex]}}.

\bibitem{ThetaCorrections}
J.~R. Ellis and M.~K. Gaillard, ``{Strong and Weak CP Violation},''
  \href{http://dx.doi.org/10.1016/0550-3213(79)90297-9}{{\em Nucl. Phys. B}
  {\bfseries 150} (1979) 141--162}.

\bibitem{dvali2022strong}
G.~Dvali, ``{Strong-$CP$ with and without gravity},''
  \href{http://arxiv.org/abs/2209.14219}{{\ttfamily arXiv:2209.14219
  [hep-ph]}}.

\bibitem{Dvali:2013eja}
G.~Dvali and C.~Gomez, ``{Quantum Compositeness of Gravity: Black Holes, AdS
  and Inflation},'' \href{http://dx.doi.org/10.1088/1475-7516/2014/01/023}{{\em
  JCAP} {\bfseries 01} (2014) 023},
  \href{http://arxiv.org/abs/1312.4795}{{\ttfamily arXiv:1312.4795 [hep-th]}}.

\bibitem{Dvali:2018jhn}
G.~Dvali, C.~Gomez, and S.~Zell, ``{Quantum Breaking Bound on de Sitter and
  Swampland},'' \href{http://dx.doi.org/10.1002/prop.201800094}{{\em Fortsch.
  Phys.} {\bfseries 67} no.~1-2, (2019) 1800094},
  \href{http://arxiv.org/abs/1810.11002}{{\ttfamily arXiv:1810.11002
  [hep-th]}}.

\bibitem{Dvali:2020etd}
G.~Dvali, ``{$S$-Matrix and Anomaly of de Sitter},''
  \href{http://dx.doi.org/10.3390/sym13010003}{{\em Symmetry} {\bfseries 13}
  no.~1, (2020) 3}, \href{http://arxiv.org/abs/2012.02133}{{\ttfamily
  arXiv:2012.02133 [hep-th]}}.

\bibitem{Dvali:2018dce}
G.~Dvali, C.~Gomez, and S.~Zell, ``{A Proof of the Axion?},''
  \href{http://arxiv.org/abs/1811.03079}{{\ttfamily arXiv:1811.03079
  [hep-th]}}.

\bibitem{PhysRevLett.53.535}
C.~Vafa and E.~Witten, ``{Parity Conservation in QCD},''
  \href{http://dx.doi.org/10.1103/PhysRevLett.53.535}{{\em Phys. Rev. Lett.}
  {\bfseries 53} (1984) 535}.

\bibitem{PQMechanism}
R.~D. Peccei and H.~R. Quinn, ``{Constraints Imposed by CP Conservation in the
  Presence of Instantons},''
  \href{http://dx.doi.org/10.1103/PhysRevD.16.1791}{{\em Phys. Rev. D}
  {\bfseries 16} (1977) 1791--1797}.

\bibitem{PQMechanism2}
R.~D. Peccei and H.~R. Quinn, ``Constraints imposed by $\mathrm{CP}$
  conservation in the presence of pseudoparticles,''
  \href{http://dx.doi.org/10.1103/PhysRevD.16.1791}{{\em Phys. Rev. D}
  {\bfseries 16} (Sep, 1977) 1791--1797}.
  \url{https://link.aps.org/doi/10.1103/PhysRevD.16.1791}.

\bibitem{WeinbergAxion}
S.~Weinberg, ``{A New Light Boson?},''
  \href{http://dx.doi.org/10.1103/PhysRevLett.40.223}{{\em Phys. Rev. Lett.}
  {\bfseries 40} (1978) 223--226}.

\bibitem{Wilczek:1977pj}
F.~Wilczek, ``{Problem of Strong $P$ and $T$ Invariance in the Presence of
  Instantons},''
\href{http://dx.doi.org/10.1103/PhysRevLett.40.279}{{\em Phys. Rev. Lett.}
  {\bfseries 40} (1978) 279--282}.

\bibitem{dvali2005threeform}
G.~Dvali, ``{Three-form gauging of axion symmetries and gravity},''
  \href{http://arxiv.org/abs/hep-th/0507215}{{\ttfamily arXiv:hep-th/0507215}}.

\bibitem{Dvali:2007iv}
G.~Dvali and G.~R. Farrar, ``{Strong CP Problem with 10**32 Standard Model
  Copies},'' \href{http://dx.doi.org/10.1103/PhysRevLett.101.011801}{{\em Phys.
  Rev. Lett.} {\bfseries 101} (2008) 011801},
  \href{http://arxiv.org/abs/0712.3170}{{\ttfamily arXiv:0712.3170 [hep-th]}}.

\bibitem{Dvali:2009fw}
G.~Dvali, I.~Sawicki, and A.~Vikman, ``{Dark Matter via Many Copies of the
  Standard Model},''
  \href{http://dx.doi.org/10.1088/1475-7516/2009/08/009}{{\em JCAP} {\bfseries
  0908} (2009) 009},
\href{http://arxiv.org/abs/0903.0660}{{\ttfamily arXiv:0903.0660 [hep-th]}}.

\bibitem{Dvali:2009ne}
G.~Dvali and M.~Redi, ``{Phenomenology of 10$^{32}$ Dark Sectors},''
  \href{http://dx.doi.org/10.1103/PhysRevD.80.055001}{{\em Phys. Rev.}
  {\bfseries D80} (2009) 055001},
\href{http://arxiv.org/abs/0905.1709}{{\ttfamily arXiv:0905.1709 [hep-ph]}}.

\bibitem{Dvali:2007hz}
G.~Dvali, ``{Black Holes and Large N Species Solution to the Hierarchy
  Problem},'' \href{http://dx.doi.org/10.1002/prop.201000009}{{\em Fortsch.
  Phys.} {\bfseries 58} (2010) 528--536},
\href{http://arxiv.org/abs/0706.2050}{{\ttfamily arXiv:0706.2050 [hep-th]}}.

\bibitem{Dvali:2007wp}
G.~Dvali and M.~Redi, ``{Black Hole Bound on the Number of Species and Quantum
  Gravity at LHC},'' \href{http://dx.doi.org/10.1103/PhysRevD.77.045027}{{\em
  Phys. Rev.} {\bfseries D77} (2008) 045027},
\href{http://arxiv.org/abs/0710.4344}{{\ttfamily arXiv:0710.4344 [hep-th]}}.

\bibitem{Blinnikov:1982eh}
S.~I. Blinnikov and M.~Y. Khlopov, ``{ON POSSIBLE EFFECTS OF 'MIRROR'
  PARTICLES},'' {\em Sov. J. Nucl. Phys.} {\bfseries 36} (1982) 472.

\bibitem{Kolb:1985bf}
E.~W. Kolb, D.~Seckel, and M.~S. Turner, ``{The Shadow World},''
  \href{http://dx.doi.org/10.1038/314415a0}{{\em Nature} {\bfseries 314} (1985)
  415--419}.

\bibitem{PhysRevLett.54.502}
D.~J. Gross, J.~A. Harvey, E.~J. Martinec, and R.~Rohm, ``{The Heterotic
  String},'' \href{http://dx.doi.org/10.1103/PhysRevLett.54.502}{{\em Phys.
  Rev. Lett.} {\bfseries 54} (1985) 502--505}.

\bibitem{Dixon:1985jw}
L.~J. Dixon, J.~A. Harvey, C.~Vafa, and E.~Witten, ``{Strings on Orbifolds},''
  \href{http://dx.doi.org/10.1016/0550-3213(85)90593-0}{{\em Nucl. Phys. B}
  {\bfseries 261} (1985) 678--686}.

\bibitem{Lebedev:2006kn}
O.~Lebedev, H.~P. Nilles, S.~Raby, S.~Ramos-Sanchez, M.~Ratz, P.~K.~S.
  Vaudrevange, and A.~Wingerter, ``{A Mini-landscape of exact MSSM spectra in
  heterotic orbifolds},''
  \href{http://dx.doi.org/10.1016/j.physletb.2006.12.012}{{\em Phys. Lett. B}
  {\bfseries 645} (2007) 88--94},
  \href{http://arxiv.org/abs/hep-th/0611095}{{\ttfamily arXiv:hep-th/0611095}}.

\bibitem{Braun:2005ux}
V.~Braun, Y.-H. He, B.~A. Ovrut, and T.~Pantev, ``{A Heterotic standard
  model},'' \href{http://dx.doi.org/10.1016/j.physletb.2005.05.007}{{\em Phys.
  Lett. B} {\bfseries 618} (2005) 252--258},
  \href{http://arxiv.org/abs/hep-th/0501070}{{\ttfamily arXiv:hep-th/0501070}}.

\bibitem{Dvali:2020wqi}
G.~Dvali, ``{Entropy Bound and Unitarity of Scattering Amplitudes},''
  \href{http://dx.doi.org/10.1007/JHEP03(2021)126}{{\em JHEP} {\bfseries 03}
  (2021) 126}, \href{http://arxiv.org/abs/2003.05546}{{\ttfamily
  arXiv:2003.05546 [hep-th]}}.

\bibitem{Halverson:2020xpg}
J.~Halverson, C.~Long, A.~Maiti, B.~Nelson, and G.~Salinas, ``{Gravitational
  waves from dark Yang-Mills sectors},''
  \href{http://dx.doi.org/10.1007/JHEP05(2021)154}{{\em JHEP} {\bfseries 05}
  (2021) 154}, \href{http://arxiv.org/abs/2012.04071}{{\ttfamily
  arXiv:2012.04071 [hep-ph]}}.

\bibitem{PhysRevD.104.035005}
W.-C. Huang, M.~Reichert, F.~Sannino, and Z.-W. Wang, ``{Testing the dark SU(N)
  Yang-Mills theory confined landscape: From the lattice to gravitational
  waves},'' \href{http://dx.doi.org/10.1103/PhysRevD.104.035005}{{\em Phys.
  Rev. D} {\bfseries 104} no.~3, (2021) 035005},
  \href{http://arxiv.org/abs/2012.11614}{{\ttfamily arXiv:2012.11614
  [hep-ph]}}.

\bibitem{Dvali:2008fd}
G.~Dvali, ``{Nature of Microscopic Black Holes and Gravity in Theories with
  Particle Species},'' \href{http://dx.doi.org/10.1142/S0217751X10048895}{{\em
  Int. J. Mod. Phys. A} {\bfseries 25} (2010) 602--615},
  \href{http://arxiv.org/abs/0806.3801}{{\ttfamily arXiv:0806.3801 [hep-th]}}.

\bibitem{Ettengruber:2022pxf}
M.~Ettengruber, ``{Neutrino physics in TeV scale gravity theories},''
  \href{http://dx.doi.org/10.1103/PhysRevD.106.055028}{{\em Phys. Rev. D}
  {\bfseries 106} no.~5, (2022) 055028},
  \href{http://arxiv.org/abs/2206.00034}{{\ttfamily arXiv:2206.00034
  [hep-ph]}}.

\bibitem{MicroBlackHolesDemocraticTransition}
G.~Dvali and O.~Pujolas, ``{Micro Black Holes and the Democratic Transition},''
  \href{http://dx.doi.org/10.1103/PhysRevD.79.064032}{{\em Phys. Rev. D}
  {\bfseries 79} (2009) 064032},
  \href{http://arxiv.org/abs/0812.3442}{{\ttfamily arXiv:0812.3442 [hep-th]}}.

\bibitem{CompactDarkMatterObjectsViaN}
G.~Dvali, E.~Koutsangelas, and F.~Kuhnel, ``{Compact Dark Matter Objects via
  $N$ Dark Sectors},''
  \href{http://dx.doi.org/10.1103/PhysRevD.101.083533}{{\em Phys. Rev. D}
  {\bfseries 101} (2020) 083533},
  \href{http://arxiv.org/abs/1911.13281}{{\ttfamily arXiv:1911.13281
  [astro-ph.CO]}}.

\bibitem{MisalignmentPRESKILL}
J.~Preskill, M.~B. Wise, and F.~Wilczek, ``{Cosmology of the Invisible
  Axion},'' \href{http://dx.doi.org/10.1016/0370-2693(83)90637-8}{{\em Phys.
  Lett. B} {\bfseries 120} (1983) 127--132}.

\bibitem{MisalignmentDINE}
M.~Dine and W.~Fischler, ``{The Not So Harmless Axion},''
  \href{http://dx.doi.org/10.1016/0370-2693(83)90639-1}{{\em Phys. Lett. B}
  {\bfseries 120} (1983) 137--141}.

\bibitem{MisalignmentAbbott}
L.~F. Abbott and P.~Sikivie, ``{A Cosmological Bound on the Invisible Axion},''
  \href{http://dx.doi.org/10.1016/0370-2693(83)90638-X}{{\em Phys. Lett. B}
  {\bfseries 120} (1983) 133--136}.

\bibitem{Dvali:1995ce}
G.~R. Dvali, ``{Removing the cosmological bound on the axion scale},''
  \href{http://arxiv.org/abs/hep-ph/9505253}{{\ttfamily arXiv:hep-ph/9505253}}.

\bibitem{Arvanitaki:2009fg}
A.~Arvanitaki, S.~Dimopoulos, S.~Dubovsky, N.~Kaloper, and J.~March-Russell,
  ``{String Axiverse},''
  \href{http://dx.doi.org/10.1103/PhysRevD.81.123530}{{\em Phys. Rev. D}
  {\bfseries 81} (2010) 123530},
  \href{http://arxiv.org/abs/0905.4720}{{\ttfamily arXiv:0905.4720 [hep-th]}}.

\bibitem{Mack:2009hs}
K.~J. Mack and P.~J. Steinhardt, ``{Cosmological Problems with Multiple
  Axion-like Fields},''
  \href{http://dx.doi.org/10.1088/1475-7516/2011/05/001}{{\em JCAP} {\bfseries
  05} (2011) 001}, \href{http://arxiv.org/abs/0911.0418}{{\ttfamily
  arXiv:0911.0418 [astro-ph.CO]}}.

\bibitem{Dvali:2021jto}
G.~Dvali, ``{Bounds on quantum information storage and retrieval},''
  \href{http://dx.doi.org/10.1098/rsta.2021.0071}{{\em Phil. Trans. A. Math.
  Phys. Eng. Sci.} {\bfseries 380} no.~2216, (2021) 20210071},
  \href{http://arxiv.org/abs/2107.10616}{{\ttfamily arXiv:2107.10616
  [hep-th]}}.

\bibitem{Dvali:2023xfz}
G.~Dvali, ``{Saturon Dark Matter},''
  \href{http://arxiv.org/abs/2302.08353}{{\ttfamily arXiv:2302.08353
  [hep-ph]}}.

\bibitem{DFSZ1}
A.~R. Zhitnitsky, ``{On Possible Suppression of the Axion Hadron Interactions.
  (In Russian)},'' {\em Sov. J. Nucl. Phys.} {\bfseries 31} (1980) 260.

\bibitem{DFSZ2}
M.~Dine, W.~Fischler, and M.~Srednicki, ``{A Simple Solution to the Strong CP
  Problem with a Harmless Axion},''
  \href{http://dx.doi.org/10.1016/0370-2693(81)90590-6}{{\em Phys. Lett. B}
  {\bfseries 104} (1981) 199--202}.

\bibitem{KSVZ1}
J.~E. Kim, ``{Weak Interaction Singlet and Strong CP Invariance},''
  \href{http://dx.doi.org/10.1103/PhysRevLett.43.103}{{\em Phys. Rev. Lett.}
  {\bfseries 43} (1979) 103}.

\bibitem{KSVZ2}
M.~A. Shifman, A.~I. Vainshtein, and V.~I. Zakharov, ``{Can Confinement Ensure
  Natural CP Invariance of Strong Interactions?},''
  \href{http://dx.doi.org/10.1016/0550-3213(80)90209-6}{{\em Nucl. Phys. B}
  {\bfseries 166} (1980) 493--506}.

\bibitem{DiLuzio:2017pfr}
L.~Di~Luzio, F.~Mescia, and E.~Nardi, ``{Window for preferred axion models},''
  \href{http://dx.doi.org/10.1103/PhysRevD.96.075003}{{\em Phys. Rev. D}
  {\bfseries 96} no.~7, (2017) 075003},
  \href{http://arxiv.org/abs/1705.05370}{{\ttfamily arXiv:1705.05370
  [hep-ph]}}.

\bibitem{Plakkot:2021xyx}
V.~Plakkot and S.~Hoof, ``{Anomaly ratio distributions of hadronic axion models
  with multiple heavy quarks},''
  \href{http://dx.doi.org/10.1103/PhysRevD.104.075017}{{\em Phys. Rev. D}
  {\bfseries 104} no.~7, (2021) 075017},
  \href{http://arxiv.org/abs/2107.12378}{{\ttfamily arXiv:2107.12378
  [hep-ph]}}.

\bibitem{DFSZ}
J.~Diehl and E.~Koutsangelas, ``{Dine-Fischler-Srednicki-Zhitnitsky-type axions
  and where to find them},''
  \href{http://dx.doi.org/10.1103/PhysRevD.107.095020}{{\em Phys. Rev. D}
  {\bfseries 107} no.~9, (2023) 095020},
  \href{http://arxiv.org/abs/2302.04667}{{\ttfamily arXiv:2302.04667
  [hep-ph]}}.

\bibitem{PhysRevD.105.085020}
O.~Sakhelashvili, ``{Consistency of the dual formulation of axion solutions to
  the strong CP problem},''
  \href{http://dx.doi.org/10.1103/PhysRevD.105.085020}{{\em Phys. Rev. D}
  {\bfseries 105} no.~8, (2022) 085020},
  \href{http://arxiv.org/abs/2110.03386}{{\ttfamily arXiv:2110.03386
  [hep-th]}}.

\bibitem{FOOT199167}
R.~Foot, H.~Lew, and R.~R. Volkas, ``{A Model with fundamental improper
  space-time symmetries},''
  \href{http://dx.doi.org/10.1016/0370-2693(91)91013-L}{{\em Phys. Lett. B}
  {\bfseries 272} (1991) 67--70}.

\bibitem{PhysRevD.17.2717}
C.~G. Callan, Jr., R.~F. Dashen, and D.~J. Gross, ``{Toward a Theory of the
  Strong Interactions},''
  \href{http://dx.doi.org/10.1103/PhysRevD.17.2717}{{\em Phys. Rev. D}
  {\bfseries 17} (1978) 2717}.

\bibitem{QCDINstantonsFiniteTemp}
D.~J. Gross, R.~D. Pisarski, and L.~G. Yaffe, ``{QCD and Instantons at Finite
  Temperature},'' \href{http://dx.doi.org/10.1103/RevModPhys.53.43}{{\em Rev.
  Mod. Phys.} {\bfseries 53} (1981) 43}.

\bibitem{Fox:2004kb}
P.~{Fox}, A.~{Pierce}, and S.~{Thomas}, ``{Probing a QCD String Axion with
  Precision Cosmological Measurements},''
  \href{http://dx.doi.org/10.48550/arXiv.hep-th/0409059}{{\em arXiv e-prints}
  (Sept., 2004) hep--th/0409059},
  \href{http://arxiv.org/abs/hep-th/0409059}{{\ttfamily arXiv:hep-th/0409059
  [astro-ph]}}.

\bibitem{Aghanim:2018eyx}
{\bfseries Planck} Collaboration, N.~Aghanim {\em et~al.}, ``{Planck 2018
  results. VI. Cosmological parameters},''
  \href{http://dx.doi.org/10.1051/0004-6361/201833910}{{\em Astron. Astrophys.}
  {\bfseries 641} (2020) A6}, \href{http://arxiv.org/abs/1807.06209}{{\ttfamily
  arXiv:1807.06209 [astro-ph.CO]}}. [Erratum: Astron.Astrophys. 652, C4
  (2021)].

\bibitem{Raffelt:2006cw}
G.~G. Raffelt, ``{Astrophysical axion bounds},''
  \href{http://dx.doi.org/10.1007/978-3-540-73518-2_3}{{\em Lect. Notes Phys.}
  {\bfseries 741} (2008) 51--71},
  \href{http://arxiv.org/abs/hep-ph/0611350}{{\ttfamily arXiv:hep-ph/0611350}}.

\bibitem{Takahashi:2018tdu}
F.~Takahashi, W.~Yin, and A.~H. Guth, ``{QCD axion window and low-scale
  inflation},'' \href{http://dx.doi.org/10.1103/PhysRevD.98.015042}{{\em Phys.
  Rev. D} {\bfseries 98} no.~1, (2018) 015042},
  \href{http://arxiv.org/abs/1805.08763}{{\ttfamily arXiv:1805.08763
  [hep-ph]}}.

\bibitem{Koutsangelas:2022lte}
E.~Koutsangelas, ``{Removing the cosmological bound on the axion scale in the
  Kim-Shifman-Vainshtein-Zakharov and Dine-Fischler-Srednicki-Zhitnitsky
  models},'' \href{http://dx.doi.org/10.1103/PhysRevD.107.095009}{{\em Phys.
  Rev. D} {\bfseries 107} no.~9, (2023) 095009},
  \href{http://arxiv.org/abs/2212.07822}{{\ttfamily arXiv:2212.07822
  [hep-ph]}}.

\bibitem{Kaplan:2015fuy}
D.~E. Kaplan and R.~Rattazzi, ``{Large field excursions and approximate
  discrete symmetries from a clockwork axion},''
  \href{http://dx.doi.org/10.1103/PhysRevD.93.085007}{{\em Phys. Rev. D}
  {\bfseries 93} no.~8, (2016) 085007},
  \href{http://arxiv.org/abs/1511.01827}{{\ttfamily arXiv:1511.01827
  [hep-ph]}}.

\bibitem{Kobayashi:2013nva}
T.~Kobayashi, R.~Kurematsu, and F.~Takahashi, ``{Isocurvature Constraints and
  Anharmonic Effects on QCD Axion Dark Matter},''
  \href{http://dx.doi.org/10.1088/1475-7516/2013/09/032}{{\em JCAP} {\bfseries
  09} (2013) 032}, \href{http://arxiv.org/abs/1304.0922}{{\ttfamily
  arXiv:1304.0922 [hep-ph]}}.

\bibitem{TKACHEV1991289}
I.~I. Tkachev, ``{On the possibility of Bose star formation},''
  \href{http://dx.doi.org/10.1016/0370-2693(91)90330-S}{{\em Phys. Lett. B}
  {\bfseries 261} (1991) 289--293}.

\bibitem{Zhang:2018slz}
H.~Zhang, ``{Axion Stars},'' \href{http://dx.doi.org/10.3390/sym12010025}{{\em
  Symmetry} {\bfseries 12} no.~1, (2019) 25},
  \href{http://arxiv.org/abs/1810.11473}{{\ttfamily arXiv:1810.11473
  [hep-ph]}}.

\bibitem{Chavanis_2011}
P.-H. Chavanis, ``{Mass-radius relation of Newtonian self-gravitating
  Bose-Einstein condensates with short-range interactions: I. Analytical
  results},'' \href{http://dx.doi.org/10.1103/PhysRevD.84.043531}{{\em Phys.
  Rev. D} {\bfseries 84} (2011) 043531},
  \href{http://arxiv.org/abs/1103.2050}{{\ttfamily arXiv:1103.2050
  [astro-ph.CO]}}.

\bibitem{Berezhiani:2021rjs}
L.~Berezhiani, G.~Cintia, and M.~Warkentin, ``{Core fragmentation in simplest
  superfluid dark matter scenario},''
  \href{http://dx.doi.org/10.1016/j.physletb.2021.136422}{{\em Phys. Lett. B}
  {\bfseries 819} (2021) 136422},
  \href{http://arxiv.org/abs/2101.08117}{{\ttfamily arXiv:2101.08117
  [astro-ph.CO]}}.

\bibitem{Berezhiani:2022buv}
L.~Berezhiani, G.~Cintia, and J.~Khoury, ``{Thermalization, fragmentation, and
  tidal disruption: The complex galactic dynamics of dark matter
  superfluidity},'' \href{http://dx.doi.org/10.1103/PhysRevD.107.123010}{{\em
  Phys. Rev. D} {\bfseries 107} no.~12, (2023) 123010},
  \href{http://arxiv.org/abs/2212.10577}{{\ttfamily arXiv:2212.10577
  [astro-ph.CO]}}.

\bibitem{Foot:2014mia}
R.~Foot, ``{Mirror dark matter: Cosmology, galaxy structure and direct
  detection},'' \href{http://dx.doi.org/10.1142/S0217751X14300130}{{\em Int. J.
  Mod. Phys. A} {\bfseries 29} (2014) 1430013},
  \href{http://arxiv.org/abs/1401.3965}{{\ttfamily arXiv:1401.3965
  [astro-ph.CO]}}.

\bibitem{AxionLimits}
C.~O'Hare, ``cajohare/axionlimits: Axionlimits.''
  \url{https://cajohare.github.io/AxionLimits/}, July, 2020.

\bibitem{Dvali:2019ewm}
G.~Dvali, E.~Koutsangelas, and F.~Kuhnel, ``{Compact Dark Matter Objects via
  $N$ Dark Sectors},''
  \href{http://dx.doi.org/10.1103/PhysRevD.101.083533}{{\em Phys. Rev. D}
  {\bfseries 101} (2020) 083533},
  \href{http://arxiv.org/abs/1911.13281}{{\ttfamily arXiv:1911.13281
  [astro-ph.CO]}}.

\bibitem{Ratra:1987rm}
B.~Ratra and P.~J.~E. Peebles, ``{Cosmological Consequences of a Rolling
  Homogeneous Scalar Field},''
  \href{http://dx.doi.org/10.1103/PhysRevD.37.3406}{{\em Phys. Rev. D}
  {\bfseries 37} (1988) 3406}.

\bibitem{Wetterich:1987fm}
C.~Wetterich, ``{Cosmology and the Fate of Dilatation Symmetry},''
  \href{http://dx.doi.org/10.1016/0550-3213(88)90193-9}{{\em Nucl. Phys. B}
  {\bfseries 302} (1988) 668--696},
  \href{http://arxiv.org/abs/1711.03844}{{\ttfamily arXiv:1711.03844
  [hep-th]}}.

\end{thebibliography}\endgroup

\end{document}